\documentclass[12pt,a4paper,onecolumn,pra,nofootinbib,
preprint,amsmath,amssymb]{revtex4}

\usepackage[french,german,british]{babel}
\usepackage[dvipdf]{graphicx}
\usepackage[a4paper]{geometry}
\usepackage{bm}
\usepackage{url}

\allowdisplaybreaks[1]

\begin{document}

\title{Understanding the special theory of relativity}

\author{\firstname{Anders} \surname{M{\aa}nsson}}
\email{andersmansson77@hotmail.com}

\affiliation{Malm\"{o}, Sweden}

\date{2009-01-29}

\begin{abstract}
This paper constitutes a background to the paper
\textit{Quantum mechanics as "space-time statistical 
mechanics"?}, arXiv:quant-ph/0501133, presented previously 
by the author. But it is also a 
free-standing and self-contained paper. 
The purpose of this paper is to give the reader an
increased and a deeper understanding of the special 
theory of relativity, and the spacetime ideas lying 
behind the above mentioned paper. We will here consider, 
discuss, define, analyse, and explain things such as, e.g., 
the constancy of the speed of light, synchronization, 
simultaneity, absolute simultaneity, absolute space 
and time, the ether, and spacetime. Albert Einstein's 
original version of the special theory of relativity 
is fundamentally an operational theory, free from 
interpretation. But the old "Lorentzian interpretation" 
and the standard "spacetime interpretation" of the 
special theory of relativity will also be considered. 
This paper also discusses and analyses aspects of the 
philosophy of science that in my opinion are relevant 
for an understanding of the special theory of relativity. 
\end{abstract}

\maketitle

\section{Introduction}
\label{sec:introd}
The purpose of this paper is to give the reader an
increased and a deeper understanding of the special 
theory of relativity. There already exist a great 
amount of literature about the special theory of 
relativity. Some of more mathematical character, but 
which are insufficient when it comes to an 
understanding of the theory. Others are more focused 
on the understanding, but seldom manage in a 
satisfactory way to give the reader just that. In my 
opinion there is missing a more gathering work on the 
understanding of the special theory of relativity. 
But also a work that in a satisfactory way explicitly 
considers, accounts for, defines, analyses and 
explains things such as, e.g., the constancy of the 
speed of light, synchronization, simultaneity, absolute 
simultaneity, absolute space and time, the ether, and spacetime. 
These concepts and definitions play an important role 
for the understanding of the special theory of 
relativity, which is something that seldom comes to 
light in the literature on theory of relativity.\\ 

This paper also discusses and analyses aspects of the 
philosophy of science that in my opinion are relevant 
for an understanding of the special theory of 
relativity. Albert Einstein's original version of the 
special theory of relativity is fundamentally an 
operational theory, free from interpretation. But 
different interpretations of the special theory of 
relativity will also be presented. One interpretation 
is the one that came in connection with a work by H. 
A. Lorentz, \textit{before} Einstein presented the special 
theory of relativity in 1905. Another more familiar 
interpretation, but which normally is not thought of 
as an interpretation, is the one that involves the 
concept of spacetime. This one came, due to Hermann 
Minkowski, \textit{after} that Einstein put forward the 
special theory of relativity in 1905.\\

This paper is a mixture of, partly, information and 
knowledge I have obtained in connection to the 
special theory of relativity, mainly from literature; 
and, partly, own analyses and how I myself have 
understood the special theory of relativity. This paper is
the result of several years of thinking about the special
theory of relativity, while I have really been working 
on other things. It has not been 
completely possible, and I have neither felt 
it to be really meaningful to try, to label every 
thought in this paper with a reference to its source. 
But references to works used, are given 
in the end of the paper.\\ 

This paper does not require that the reader is 
previously familiar with the special theory of 
relativity, but it makes it easier for the reader if 
this is the case.\\
  
This is a free-standing and self-contained 
paper. But it also constitutes a background to the paper 
arXiv:quant-ph/0501133 presented previously by the
author.~\footnote{See also pages~39--51 of
  the author's thesis, which can be downloaded at\\
  http://www.diva-portal.org/kth/theses/abstract.xsql?dbid=4417}\\ 

I will begin by giving a background to and 
an outline of the special theory of relativity. Some 
undefined concepts will appear in this background, 
but this will be remedied as we go along.

\section{Background to and overview of the special theory
  of relativity}
\label{sec:backgr}
In the end of the 19th century, physicists tried to understand 
and unite existing physical theories with observations 
that were available at that time. There were clues that 
the existing theories needed to be revised. During the 
19th century there were more and more clues that 
light had wave properties. But wave properties were at 
this time only something that occurred in relation to a 
medium of some sort. When it, e.g., comes to sound 
waves, then they are a pressure phenomenon in air, and 
water waves are water in motion. Therefore it was 
believed that it had to be the same with light, since 
this also seemed to be a wave phenomenon.\\ 

But there were no observations of such a medium, 
or the \textit{ether} as 
it was called. The ether was thought to exist 
everywhere and to be able to penetrate through all matter. 
Its nature had many similarities with Newton's \textit{absolute 
space}, which also was not directly detectable with any 
known measurement procedures or observations. There 
were different variants of theories that described how 
this ether was supposed to work and interact with other 
matter. One tried to measure and find signs that such 
an ether existed, but the attempts turned out to be 
fruitless or unsatisfactory in one way or the other. 
E.g., if the ether was a medium for wave motion like 
any other, then one expected that the velocity of light 
relative to the source, should be dependent on the 
velocity of the source relative to the ether. But there 
were no indications that this was the case. Instead 
there were indications that the speed of light, relative 
to \textit{every} observer in constant uniform motion, always was 
the same, independent of the speed of the light source. 
These observational indications were hard to
reconcile with the ether hypothesis, without at the same 
time giving the ether more and more strange properties. 
On the whole, these observational indications were hard 
to make compatible with everyday thinking and a classical 
physical description of reality.\\    

	The physicist H. A. Lorentz lived and worked 
during this time, and that before Einstein entered the 
scene. Lorentz changed the transformation laws for how 
space and time coordinates change when one changes 
\textit{inertial reference frame}, i.e., reference frames or 
reference bodies that are in constant uniform motion, to 
make these consistent with the laws of electromagnetism. 
As a matter of fact, Lorentz arrived at \textit{exactly the same} 
equations, the so-called \textit{Lorentz equations}, as Einstein 
later also did in his special theory of relativity. The 
difference between their presentations was in the way 
they had come to and interpreted these equations. The 
Lorentz equations describe how space and time must change 
for bodies in constant uniform motion. It follows from 
these equations that lengths are shortened and that time 
goes slower for bodies in motion.\\

The most fundamental difference between Lorentz's 
interpretation and Einstein's interpretation of these 
equations, was that Lorentz assumed the existence of an 
underlying absolute Newtonian space, whereas Einstein did 
not do this. According to Lorentz's interpretation, there 
are bodies that really, truly, or objectively seen, are 
in \textit{absolute rest} relative to the absolute Newtonian space; 
or relative to the ether, which in many respects plays a 
similar role to that of the absolute Newtonian space. The 
"Lorentzian interpretation" means that for all bodies 
which move relative to the absolute Newtonian space, time 
\textit{objectively seen} goes slower and lengths 
\textit{objectively seen} become 
shorter than for bodies which are in absolute rest. 

However, the Lorentz equations at the same time mean
and enable that constant uniform motion relative to the absolute 
space cannot be detected by trying to observe such changes in 
lengths and time. The only effects which can be observed, 
are only dependent on the relative velocity between 
reference bodies. All reference bodies think that it is 
the time on other reference bodies that goes slowly and 
lengths are shortened. Below we will come back to how this 
"illusion" is possible according to such an interpretation. 
For bodies in constant uniform motion relative to one 
another, the situation is thus completely symmetric and the 
same laws of nature hold for all inertial reference
systems. This is in line with the \textit{Galilean principle of 
relativity}, because ever since Galilei it has been known 
that the classical laws of physics are the same for every 
observer in a state of constant uniform motion. Only the 
relative velocity can be measured, and not who "really is 
in motion" or who "really is at rest".\\

As time went on, it became more and more clear to Einstein 
that the ether hypothesis seemed to be superfluous and perhaps 
even incorrect. Einstein was strongly inspired and 
influenced by the physicist Ernst Mach, but also by others, 
e.g., the philosopher David Hume. One can perhaps say that 
Einstein's approach to the problems concerning space, time 
and the nature of light, was an application of Mach's 
philosophical ideas and view on science. As Mach did, 
Einstein also tried to explain physical phenomena without 
resorting to hypotheses that did not have support in 
observations. Newton's absolute space, and the ether, were 
to Einstein hypotheses that did not have any direct support 
in observations, so perhaps they did not exist? In the 
spirit of Mach, Einstein assumed that there was no 
underlying ether, or absolute Newtonian space. Einstein 
generalised the Galilean principle of relativity to include 
Maxwell's theory of electromagnetism.\\

However, strictly seen there can only be motion 
relative to other bodies, if one only wants to define the 
concept of motion based on what one can observe. But if all 
motion is relative, what does one really mean when one says 
that a reference body is in a "state of motion" or in 
"constant uniform motion", if one does not at the same time 
specify relative to what the reference body moves?\\ 

Einstein was completely clear about and aware of these 
problems. In spite of this, he kept in the special theory 
of relativity the idea that bodies can be in a state of 
constant uniform motion, without one having to specify 
something that they should be in motion relative to. The 
special theory of relativity is limited to hold only for 
inertial reference systems. In a way this means that 
Einstein kept some sort of absolute space concept after all. 
Thus, in the special theory of relativity, the concept of 
\textit{absolute motion} is still there.\\ 

However, motion is only 
absolute for bodies in accelerated motion. And Einstein did 
not think of it as motion relative to an absolute Newtonian 
space, because in the special theory of relativity there is 
no such thing as an absolute Newtonian space. There are no 
privileged inertial reference systems in the special theory 
of relativity, i.e., no inertial reference system can be 
said to be in absolute rest. But the inertial reference 
systems can be said to be in a state of "absolute" or 
objective constant uniform motion. It was first with the 
general theory of relativity that there was no absolute 
motion at all. All states of motion and reference systems 
are equivalent in the general theory of relativity, no 
matter how the reference bodies are moving. The difference 
between the special and general theory of relativity is 
that the latter one includes acceleration and gravitation, 
whereas the former does not.\\ 

The special theory of relativity 
must therefore keep an element of absoluteness in the form
of constant uniform motion, i.e., inertial reference systems 
as privileged reference systems when it comes to formulating 
laws of nature. In the general theory of relativity, inertial 
reference systems are no longer privileged reference systems. 
There are no privileged reference systems at all in the 
general theory of relativity and the laws of nature are the 
same in every reference system.

\section{The Newtonian world view}
\label{sec:newtww}

An interesting question is whether the idea or notion of
the \textit{absolute Newtonian space} is something Newton invented
and since then has indoctrinated all of us with? 

Or is it perhaps something biologically given to us all 
and for that reason we experience it as right? A concept 
that Newton only refined and then could use to formulate
and make his physical ideas more exact?\\

Personally I tend to think that the second alternative is 
the right one, but that also the first alternative has had 
a great significance and influence on our thinking. The 
notion of the Newtonian space has become so natural to us 
that it would be difficult to free ourselves from it, even 
if we wanted to. I would say that this is because the 
"Newtonian approach" agree well with our everyday 
understanding of space and time.\\
 
What do I then mean by the \textit{absolute Newtonian space}? Well, 
one could think of it as a (infinitely) large empty space, 
room, or void. Like a gigantic box or aquarium without walls 
that contains a huge empty space. This space has a geometric 
and metric structure, given by forming a coordinate system 
consisting of three coordinate axes - height, length and 
breadth. The coordinate axes are perpendicular to each
other and the coordinates are usually denoted by $x$, $y$, $z$; 
i.e., a \textit{Cartesian coordinate system} (see figure~1).\\
 
\begin{figure}
\includegraphics[scale=.8]{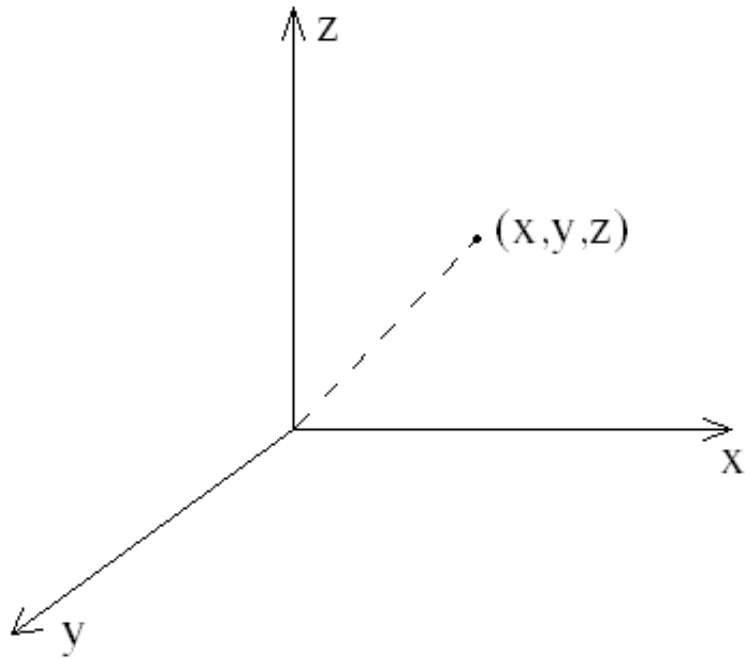}\\
\caption{A Cartesian coordinate system.}
\end{figure}

Now it is actually not the space in itself one experiences, 
but instead all the objects that fill up the space, such as 
galaxies, planets, cars, humans, atoms and light. The space 
in itself cannot be experienced directly. The existence of 
an empty space and its hypothetical effect on the objects 
filling up the space, can only be experienced indirectly
by experience the objects themselves and how they move or 
"want" to move in relation to other objects. The empty 
space in itself is something one imagines to exist.\\

One could think of the empty space as something that
exists irrespective of if there is something or not which 
"fills up the empty space". So even if one would remove 
everything which is "in the empty space", one could still 
imagine that there would remain a three-dimensional empty 
space with a geometric and metric structure that gives the 
distances between positions in the empty space. (We are 
here coming in contact with questions having to do to
with what we consider to be real or unreal, which is
something we will came back to below.)\\

However, the physical objects that are in the empty space, 
do not just stand still in space, but can also move. It is 
in particular here that time enters the picture, even if 
there is also a meaning in speaking of time, and that
one can have a subjective feeling of a time in motion,
even if one does not see any objects in one's surrounding 
which are moving. Time is, of course, measured on a clock, 
but in the Newtonian world view one imagines that there 
exists an ideal \textit{absolute time} that ticks on at the same 
rate independently of what there is, and what is happening, 
in the absolute Newtonian space. How physical objects are 
moving in space, one can describe by giving their space 
coordinates $x$, $y$, $z$, for every moment in 
time t on the absolute time.\\ 

The absolute Newtonian space is like a snapshot of the spatial 
three-dimensional space at a certain point of time on the 
absolute time. Only the three-dimensional spatial space at 
the moment of time "right now", i.e., \textit{the present} 
or \textit{now}, has an existence (see figure~2). Thus, for all 
other points in time the spatial three-dimensional space 
does not exist. Instead these snapshots of the spatial 
space belong to the past or the future, which means that 
they have either previously existed or have not yet come 
into existence, respectively. In other words, reality 
exists only "one moment at the time".\\

\begin{figure}
\includegraphics[scale=.8]{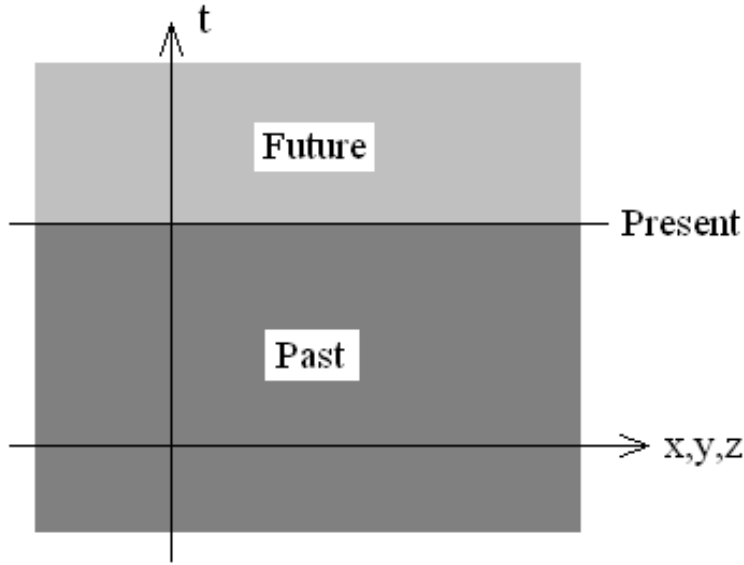}\\
\caption{The past, present, and future with an Newtonian
  world view.}
\end{figure}

There is another property that characterize the absolute 
Newtonian space. Already Galilei knew that it is not 
possible to speak about absolute motion for unaccelerated 
bodies, i.e., for bodies in constant uniform motion. For 
imagine that you are on a train which is at a train 
station and that you look out through the window. If one 
sees that a train on the neighbouring track moves, it can 
sometimes be difficult to tell whether it is the train one 
is on, or if it is the other train which is in motion 
relative to the train station, if one, e.g., does not feel 
any vibrations due to the motion of the train one is in, 
and if one does not see the ground or the train station. 
In the same way it can be hard to separate the situation 
in an airplane at an altitude of 10 000 meters flying at 
1000 km/h, from a stay in a sofa at home in one's living 
room, if it was not for the fact that one, e.g., heard the 
sound or felt the vibrations from the engines of the 
airplane and the air friction on the fuselage. Also 
remember that the Earth is moving around the sun with 
great speed, and that the solar system in its turn is 
moving around the Milky way, and so on, without us sensing it.\\ 

All the above mentioned, are examples of that only 
\textit{relative motion} can be observed when it comes to bodies in 
constant uniform motion. Therefore, when it comes to 
unaccelerated motion, one cannot tell who or what, 
"actually" or "really", is moving or is at rest. Neither 
is it possible to say who or what, is really moving or at 
rest, relative to empty space, whatever that would mean?\\

What is the situation then, when it comes to accelerated 
motion? Can one in this case tell who or what is really 
moving? Suppose that two space ships are at rest relative 
one another. One of the ships starts its engine and thus 
begins to accelerate, as the passengers experience through 
that they are pressed against the back of their seats and 
that things that lie loose, or float freely, are set in 
motion. The passengers on the other space ship, of course, 
do not feel any acceleration. Or consider a rotating body 
in empty space. A person on this body will experience a 
centrifugal force. These kind of effects can reveal if 
one's reference body accelerates or not.\\

But if there only is relative motion when the motion is 
uniform, why is there not also only relative motion when 
the motion is accelerated? Relative to what is, e.g., the 
spaceship accelerating in the above example? And relative 
to what is a planet rotating? To be able to handle these 
kinds of problems, Newton imagined that bodies that are 
accelerating or rotating, are doing this relative to an 
underlying empty space which is in absolute rest, which is 
what we call the absolute Newtonian space. However, all 
bodies in constant uniform motion relative to this 
absolute space are equivalent. I.e., reality looks the 
same and all laws of nature are identical relative to such 
reference bodies or reference systems, so-called \textit{inertial 
reference systems}. Only relative to such privileged 
reference systems, the laws of nature in classical physics 
hold.\\

In presenting his general theory of relativity, Einstein 
generalised this Galilean principle of relativity to also 
hold for accelerating reference bodies. According to this 
"generalised Galilean principle of relativity", the laws 
of nature are the same relative to all reference bodies no 
matter how they move. That Einstein was able to do this, 
is because he put an equality sign between the effects of 
acceleration and a gravitational field.

\section{The concepts of space, time, and simultaneity}
\label{sec:concsptisi}

What is actually meant by "right now" when one speaks of 
"everything that happen right now"? If the speed of light 
was infinite, then one would \textit{see} something happen in the 
same instant as it \textit{occurs}. However, it is true that the 
speed of light is very large, more precisely 300 000 000 
meter/second, but it is not infinitely large. This means 
that if one sees something happen, it does not occur at
the moment one sees it happen, but rather it occurred at 
an earlier moment. 

In the same way it is with sound, which 
also propagates with a finite speed. To hear something, 
therefore does not mean that what one hears, also happens 
at the moment one hears it, which the everyday phenomenon 
of echo clearly demonstrates.

If one, e.g., takes a photo of a mountain from a couple of 
tens of kilometers distance, then it is not how the 
mountain really looked when one took the picture, but 
strictly seen how the mountain looked a couple of ten or 
hundred thousands parts of a second before one took the 
picture. In everyday terms this is, of course, a very
short time. But on an astronomical scale there is a big 
difference between that something \textit{occurs} and 
that one \textit{sees it occurring}. The stars 
and galaxies as we see them, are 
usually how they looked several years ago. Exactly how
many years ago of course depends on how far from us they 
are. E.g., a supernova that we see in the starry sky,
could have happened when Newton lived, but on Earth we did 
not see it happen until now.\\

Therefore is, strictly seen, everything that we experience 
in our surroundings, not something that occur in the moment 
of time we experience it, but instead something that 
occurred at an earlier point in time. But when it comes to 
everyday phenomena which happen here on Earth, and I then 
primarily think of phenomena which involve light (or more 
generally electromagnetic radiation), one in practice sees 
and experiences things at the same moment as they occur. The 
time it takes for the light to go from where it was sent to 
where it is registered, as, e.g., an eye or a camera, is so 
short that for most practical purposes one can ignore it. In 
practice we can therefore say that, at the moment in time 
when the picture was taken, the mountain really looked as it 
does in the photograph.\\

Perhaps one can in this find an explanation to or support for 
the idea that absolute simultaneity and a "Newtonian world 
view" could be innate notions about reality. Since the speed 
of light in practice seems and can be taken to be infinite, 
evolution could have equipped us with a conception of reality 
that does not take into consideration the fact that the speed 
of light actually is finite. What we can \textit{see} "right now", 
could then be something that we also imagine \textit{occurs} "right 
now". Our \textit{subjective} or \textit{personal present} 
would then not be 
limited only to ourselves and our closest surroundings, but 
would include the whole space that we can see; which in principle 
could be unlimited, i.e., infinitely large.

That we always must take into consideration the fact that the 
speed of light is finite, to determine whether two events 
occur at the same time or not, is something we historically 
seen from relatively modern scientific progress are aware of, 
but which I think we dismiss as a practical and not as a 
principal problem. In other words, we imagine that it in 
principal really is meaningful to speak about the 
simultaneity of two events, despite the fact that we 
consciously or unconsciously suspect that we would probably 
get ourselves into practical problems and difficulties if we 
actually would try to determine whether two specific events 
are simultaneous or not.\\

But if what we see, does not occur at the moment we see it 
occurring, is it then really meaningful to say that two (or 
more) \textit{events} in space occur 
\textit{simultaneously}? For what does 
one really mean when one says that two events occur 
simultaneously? Concretely, what would you do, if you wanted 
to determine what happens \textit{simultaneously} somewhere else in 
the world or universe? Is it at all possible to determine 
that? In other words, is it really meaningful to wonder about 
what happens somewhere else in the world or the universe 
"right now"? By "right now" I mean exactly at that moment 
when one wonders about what happens somewhere else in the 
world or universe.\\

If one thinks that it obviously is meaningful to speak about 
the simultaneity of two events, without having to specify 
more closely what one means by this, one probably has an 
unconscious notion of \textit{absolute simultaneity} in the back of 
one's head. But absolutely simultaneous relative to what? 
Well, relative to something objective. And what would this 
objective thing be if not an absolute space, in relation to 
which events are absolutely simultaneous!? But to make it 
more clear, and hopefully also more convincing, what I say 
and mean, let us make an attempt at analysing the origin of 
the concept of \textit{absolute simultaneity}.\\

However, before doing this, let me first define what is meant 
by an \textit{event}. An \textit{event} is defined as a 
designation of \textit{where} 
(i.e., the three space coordinates relative to a reference 
body) together with \textit{when} (i.e., the point in time) something 
occurs. The three-dimensional (spatial) space we can imagine 
as a "crystal structure" or "lattice" of perpendicular 
measuring rods (with some suitable length unit). However, 
we must be more precise about what we mean by the "point
in time" when something occurs, i.e., according to which 
clock? Well, according to the clock which is at the
position in space where the event occurred. In practice 
there is, of course, no clock, and it would be impossible 
to place a clock, at \textit{every}
position in space. But in principle we can imagine it to 
be possible and that we have placed a clock at every 
position in space where the measuring rods meet in the 
"lattice" (see figure~3).\\

\begin{figure}
\includegraphics[scale=.8]{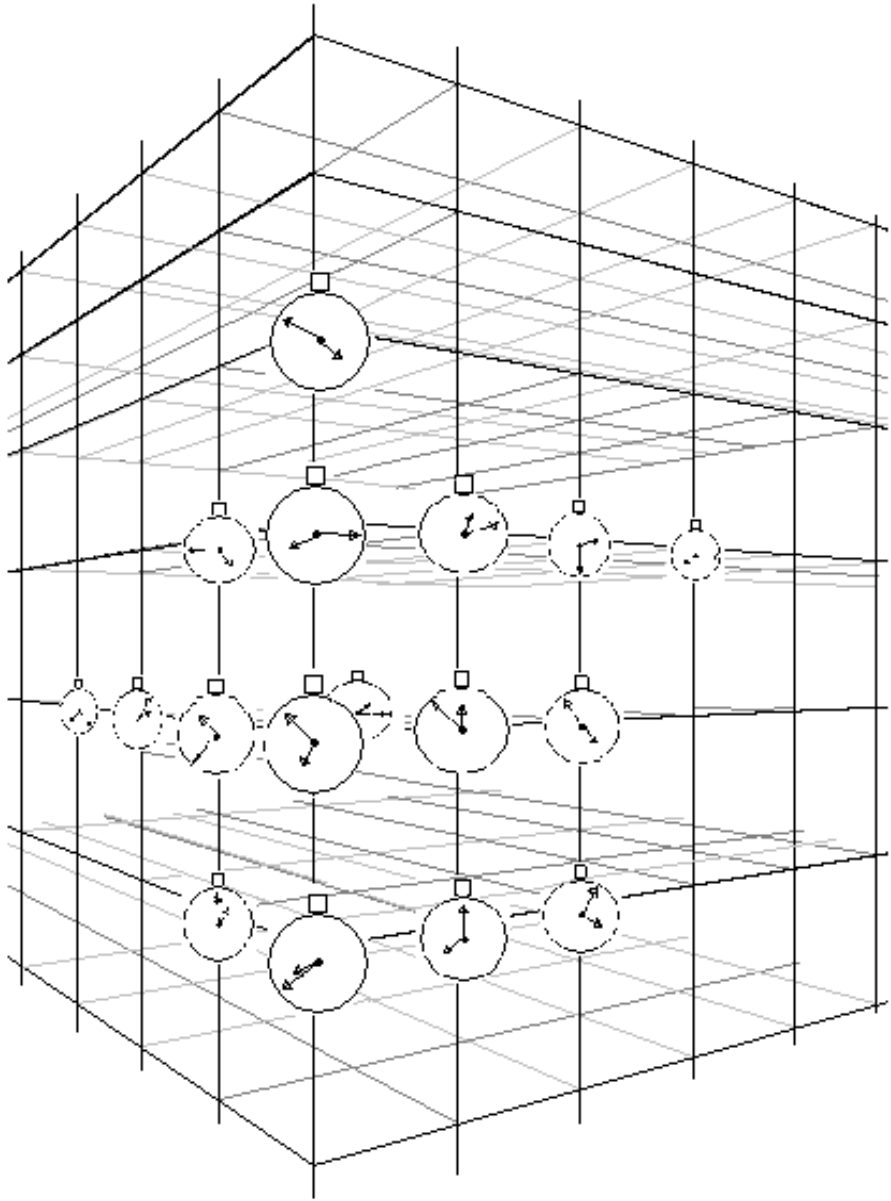}\\
\caption{An implementation of a coordinate system in space and time.}
\end{figure}

If one does not have too high thoughts about oneself or 
is not too philosophical, it is natural to believe that 
one's own existence is not more unique or more special 
than anybody else's. A reasonable assumption is that other 
people exist to the same degree and fundamentally in a 
similar way to how we experience ourselves. Let us
therefore assume that other persons experience time 
in the same way as we ourselves do. I experience the 
present, remember the past, and the future is something 
that has not yet come into existence. For me only the 
subjective or personal present is something I experience 
to exist. I exist in one and only in one moment in time, 
and that is the personal present. I therefore imagine that 
others experience time in the same way.\\ 

A person's existence can be connected with the position or 
place in space where this person is. Because, the place in 
space where I am at, e.g., the ground under my feet, in 
principal constitutes a physical extension of my own body, 
i.e., an extension of myself. I thus experience and imagine 
that the place in space where I am at, exists in the same 
way as I myself exist. Hence both I and the place where I 
am at in space, exist in the personal present. 

The personal present corresponds to a specific point in time, 
which can be read on the clock located in space where I am at. 
So instead of talking about the personal present, one 
could just as well talk about the time that a person in
his personal present reads on his clock; where by "his 
clock" is meant the clock located where the person is 
located in the three-dimensional (spatial) space. Let us 
call this point in time the \textit{local present}. The local 
present is thus the point in time corresponding to the 
personal present; or in other words, the point in time 
which the personal present constitutes. Just as the
personal present is connected to a certain person, 
the local present only has a meaning, if one also 
gives the position in space of the clock that one refers to.\\

Now there can in principal be a person at every place in 
the three-dimensional (spatial) space. One can therefore 
imagine that there is an existence of a personal present 
at every place in space. But the place in space where one 
is at, we said that one could regard as an extension of 
oneself. And the personal present can be replaced by the 
local present on the clock located where one is at. So 
instead of speaking about that "a person exists in a 
personal present at a place in space", one can therefore 
think away the person's existence and only speak about "a 
place in space that exists in the local present".\\

But how are all these instantaneous existences of places 
in space related to one another? We have imagined that 
every place in space only exists in the local present. 
The totality or general picture of all these places' 
instantaneous existences is one single connected and 
unitary snapshot of a three-dimensional spatial space.\\

In this way the notion of \textit{absolute simultaneity} and an 
\textit{absolute three-dimensional spatial space} could have
arisen. So despite the fact that we only experience the 
space in our closest vicinity or surroundings at a single 
moment in time, the personal present, we imagine that 
there exists one single (infinite) three-dimensional 
(spatial) space which exists at one single moment in 
time in common to us all - one common personal present. 
An idea of absolute simultaneity thus seems to be the 
price one has to pay if one imagines reality in the above 
described way.\\

Since the experience that we only exist in a subjective or 
personal present feels so obvious to us and is so deeply 
rooted in us, normally it does not even occur to us to 
contemplate, even less question, this experience or notion. 
But once we have become aware of this notion and how 
closely associated it is with the concept of absolute 
simultaneity and absolute space, we can also begin to 
question and try to change this conception. This could 
give us more freedom in forming concepts and thus how we 
construct our theories that describe reality. By questioning 
the concept of simultaneity Einstein made it possible for 
himself to get away from the assumptions of Newton's 
absolute space and the ether. Einstein approached the 
problems connected to the mysterious nature of light from 
a philosophical point of view that was influenced and 
inspired in particular by the physicist Ernst Mach's 
philosophical approach to science and reality. Einstein 
eventually arrived at the same equations as Lorentz did, 
but with a different interpretation of the concepts of 
space, time and simultaneity.

\section{The philosophical approach of Mach and Einstein}
\label{sec:philappmaei}

Einstein was thus clearly influenced by Mach's philosophical 
approach (but also by, among others, the philosopher David 
Hume). Mach denied in principle everything that was not 
observable or measureable. This lead, among other things, 
Mach to deny the existence of Newton's absolute space, since 
this was immeasureable according to Mach. (It is true that 
this philosophical approach also lead Mach to deny the 
existence of the atom, but this is a story in itself which
I wont go into here.) According to Mach, it was not 
meaningful to speak about the motion of a body in relation 
to an absolute space. He tried instead to explain the
inertia of a body, and the effects of acceleration which 
arise relative to an accelerating reference body, as a 
consequence of the relative motion of the reference body
in relation to all other stars, planets, etc; i.e., relative
to the fixed stars.\\ 

At first sight, this can perhaps seem 
to be a rather strange, and maybe even naive, way to
approach the problem of inertia and acceleration. But 
admittedly there is clearly an unsatisfactory element to 
base physics on concepts such as an absolute space and 
motion in relation to this space. Because the absolute space 
is hardly something that can be said to directly correspond 
to something in our sensations or observations of reality. 
E.g., one cannot see or experience a black empty 
space moving. For imagine that you are in empty space without any 
planets, stars, galaxies, etc, in sight. The black empty space 
around you looks the same regardless if you are at rest, are 
in uniform motion, accelerating or rotating. Regardless of 
your state of motion it is the same black empty space you 
experience and see before you. Newton's absolute space is 
thus hardly something that can be said to correspond to 
something in our sensations or observations of reality. It 
only exists in our imagination and in our theories.\\

According to Einstein's and Mach's philosophical approach, 
we are not allowed - if we are to be consistent - to assume 
the existence of something that we cannot observe. There is 
something appealing, not to say obvious, to try to base physics 
only on things that find support in observations. 

For has not the cast of roles been inverted, if it was to be us who 
should tell reality, which notions and concepts that should 
have a counterpart or a correspondence in reality, and not 
the other way around? Is it really our task to tell reality 
which concepts it should contain, and what should be real or 
not? For should it rather not be reality that should tell us 
what it is that exist and do not exist? Instead of trying 
to describe and explain reality on the basis of concepts we 
ourselves have created, should we not describe and explain 
reality only with concepts whose meaning is defined based
on, and do not exceed or go beyond, what we can observe
and measure? 

But then, what about concepts and notions we have acquired or 
been equipped with through evolution? Because when the 
human body with its brain was formed through 
evolution, must not the concepts have 
been formed based on what was available for and could
affect the constituents and building blocks of our brain? 
Or how could it otherwise have been or come about, seen 
from an evolutionary perspective?~\footnote{Also our logic or
logical reasoning is reasonably the result of
  evolution, and have been formed by how reality is and
  functions. But since logic is intimately connected with how we
  reason, our logic itself is more complicated (and
  probably partly impossible) to analyse and question,
  since we need logical reasoning to reason about logical 
  reasoning. However, the aim here is not to get too 
  involved or go too deeply into these sort of questions, 
  so I will not go further into this here.}\\

If we speculate that in the brain there was created a 
notion of absolute simultaneity and an absolute space, 
did the brain then not go beyond experience and observations 
of reality? Possibly, but it is not difficult to come up 
with a possible explanation as to why the brain could have 
done something like that. E.g., it perhaps made the world 
around it easier to grasp and thereby gave the body, that 
the brain was in, an evolutionary advantage. And us 
humans are not necessarily constructed through evolution
to be able to experience and understand reality as it 
really is, but only to understand and control the everyday 
world in which our bodies for all practical purposes 
live in.\\

The hypotheses of the existence of the atom, or
strings in string theory, are also examples of concepts that go 
beyond experience and what can be observered, at least 
when the concepts were first invented. But in spite of 
this, they are examples of concepts that have turned out 
to be successful. At least the atom, which is a hypothesis 
that has been strengthened more and more as physics has 
progressed. It should not be very controversial to say, 
that the hypothesis of the existence of the atom, has been 
accepted and in practice has been taken to be correct by a 
majority of scientists, as well as the rest of
society. Then it is possible that atoms will never be felt 
or thought to exist in the same way as a football is 
experienced to exist. The atom is an example of a 
hypothesis which is not directly, but rather indirectly, 
based on observations. It is a concept which was created 
in order to be able to successfully understand, describe 
and explain observations. But as things progressed, the 
atomic hypothesis was supported, strengthend and confirmed 
by more and more experiments and theories. The atom 
hypothesis was finally accepted and atoms thought to be 
something real. Therefore one cannot reject or discard a 
concept or notion just because it does not find support 
in today's observations, knowledge and theories of reality.\\

What was described above, one could describe as two different 
ways of working or philosophies of science. Historically seen 
science has successfully used both of these approaches. There 
seems to me, to be a mutual interplay or symbiosis between 
these two approaches. When it comes to the situation which 
was before Einstein and other physicists in the years that 
preceded the year of 1905, when Einstein presented his special 
theory of relativity, then with the result at hand, Mach's
and Einstein's approach seems to have been the right one. But 
the ether hypothesis could just as well have turned out to
be the right one. 

Furthermore, is it not so that, what one 
takes as right or wrong often are after constructions? 
Because often it is only when one knows the result of 
something, that one with certainty can say what is right or 
wrong. Is it not so that, what really decides if we take 
something to be right or wrong, a matter of how successful 
this something turns out to be? And if something is 
successful or not, is strictly seen something that only can 
be judged afterwards, with the result at hand.

\section{What is real?}
\label{sec:whatisreal}
 
When one, as here, discusses philosophical and fundamental 
questions in physics, one easily comes into questions concerning 
what one is to consider as real and unreal. It is natural 
and one can hardly avoid to consider something as real if it 
directly affects our senses, as, e.g., the sun, a table, a 
glass of water, or an apple falling from a tree. (But if one 
wants to play the devil's advocate for a moment and point out, 
or rather state, that one from a philosophical standpoint 
could argue that nothing is real. But on the person arguing for 
something radical like that, there is also the obligation on 
him or her to explain what he or she really means by such a statement. 
Then one could also ask oneself how successful or meaningful 
such an approach is when it comes to describe, predict, 
understand and explain what we experience and observe.)\\ 

When it comes to such things as, e.g., air, heat, smells, etc, it 
is hard to not also consider these as real, since they all
can be observered with one or more of our five senses (i.e., 
eyesight, hearing, smell, taste, and touch). But I would also 
say that air is not something we experience as real in the
same way or to the same degree as we experience, e.g., a 
table as real. A table we can see, touch and feel, but also 
hear, taste and smell. It can affect all our senses. It is, 
e.g., also a more solid object, with a particular shape
and position in the room, and it can clearly be separated 
from its environment. The air, on the other hand, we
cannot see, taste or smell, but we can feel it and hear
it. It is, e.g., more ephemeral and does not really have 
any shape, and it is not clearly separated from its 
environment as, e.g., a table is.\\

We can go one step further and also consider, e.g., gravitation, 
magnetic fields and different kind of forces as real. But 
why and with what right do we do this? One possible answer 
is that one often considers such, not directly observable, 
things as real, when they give rise to effects that can be 
distinguished or separated from "the natural order of
things", by which is meant how things usually are and 
behave under normal circumstances. In the time before us 
humans freed ourselves from our directly earthbound life 
and came to look (more closely) at the stars, it probably 
appeared as obvious and natural that bodies always fall 
towards the ground. With the limited knowledge of the 
world one then had, it would not have been strange if one 
then did think that it was not needed anything real to 
\textit{cause} things to fall towards the ground.\\ 

But with astronomy and the idea that Earth and the Sun are not 
the centers of the universe, and that bodies do not 
always fall towards the ground, one started to question 
one's previous view and knowledge of reality. The natural 
order of things instead became that a body moves uniformly 
in a straight line, until it is acted on by forces, such
as in a collision, or by frictional or gravitational
forces. Forces could then be viewed as real since they 
constituted a deviation from how things normally behave. 
With generally applicable laws and mathematical equations, 
Newton could with precision describe and explain the motion 
of bodies, planets, and stars, as a consequence of action 
of forces. It thus became convincing and natural to consider 
gravitation and gravitational forces as something real. 
However, nothing forces us to view gravitation or 
gravitational forces as something real, since they do not 
affect our senses directly. We observe gravitation only 
indirectly through its \textit{thought} or 
\textit{hypothetical} effect on 
something that does directly affect our senses and which
we therefore can perceive. 

Consider, e.g., an apple falling 
towards the ground. We do not here perceive, experience or 
observe gravity in itself, but only the apple and its motion 
relative to other objects. Gravity is something we imagine 
exists and that it pulls the apple towards the ground. It is 
true that we can feel that something pulls us towards the 
ground. But if I, e.g., feel that gravity pulls my hand 
towards the ground, I do not really feel gravity itself, 
but rather that the hand pulls the muscles in my wrist. 
And if one finds oneself in free fall under the influence 
of gravity, one does not experience or feel any gravitational 
forces at all.\\

Furthermore, in the classical theories of physics one does
not normally consider such things as, e.g., air, heat, 
pressure, energy, and momentum as something directly real, 
but as a \textit{composition} or 
\textit{consequence} of something else
which in turn is considered to be real. If one, e.g., 
assumes that atoms exist, then air becomes just a big 
swarm of atoms (molecules), and pressure becomes the total 
force and macroscopic effect of many atoms colliding with 
other objects. Energy and momentum are examples of 
quantities defined in terms of other quantities, such as 
mass, position, time, and velocity. The latter 
quantities are normally considered as something real, 
unlike energy and momentum which are mathematical 
compositions of quantities such as mass, position, time, 
and velocity. 

It is true that the quantity velocity is a 
mathematical composition of position and time. But since 
velocity, unlike energy and momentum, has a direct 
counterpart among our sensations, with the property "to 
be in motion", I here choose to consider velocity as 
something real; but since this does not have any real 
significance for the rest of this paper, I wont go into 
this any deeper than that.\\

Generally seen there is hence some room for interpretation 
and we have a certain degree of freedom, when it comes to 
choosing what we consider to be real or not, and
exactly what this something should be once we have decided to
consider it as real. As is 
well known, Newton explained gravitation as a force that 
acts between massive objects. However, in Einstein's 
general theory of relativity there are no gravitational 
forces. Instead gravity is there an effect of how 
\textit{spacetime} curves. 

In the case of atoms, one cannot observe 
atoms directly. (Remember that a scanning tunnelling 
microscope only gives an indirect image of atoms \textit{via} the 
theory of quantum mechanics.) But by assuming that atoms 
exist, many pieces fall into place. The multitude of 
different things that one can explain, understand and 
predict by assuming that atoms exist, are so overwhelming 
and convincing that it is hard to deny the existence of 
atoms. Since the atomic hypothesis is so successful, it is 
in the current situation rather on those who deny the 
existence of atoms to, in addition to explaining why he 
or she does not believe that atoms exist, try to find a better 
and more satisfactory underlying explanation or hypothesis 
than the atomic hypothesis. 

By this is not meant that the 
understanding and description of atoms will 
not need to be revised in the 
future. This is something that (directly or indirectly) 
happens all the time, as one comes to understand more 
and more about reality and its most fundamental 
constituents.\\
	
Before we continue, I just briefly want to explain what 
my own personal thoughts are on such words as "to exist", 
"real" and "reality". I have an \textit{ontological} approach to 
the world. I imagine that there is an objective world 
"out there", which we are a part of, but which exists 
independently of us humans. As I see it, we humans are 
just a product of that which exists objectively and is 
"out there". 

For me, the ultimate goal and ambition of 
science is, as well as we possibly can, to describe and 
understand this objective reality we all find ourselves 
in, and are a part of. I therefore imagine that some 
things are objectively real, while other things are 
only effects or compositions of things that are 
objectively real. E.g., for me fields and elementary 
particles are more natural candidates to be something 
fundamental and objectively existing, than I, e.g., 
believe that probability or energy is. I do not imagine 
probability or energy as things which have objective 
counterparts "out there". For me they are just abstract, 
mathematical quantities and concepts that exist only as 
a thought in our brains. And a brain is for me only a 
sophisticated biological machine which is a result of 
evolution, which in its smallest constituents consists 
of elementary particles and fields. Particles and fields 
(perhaps also strings, membranes, and other variants of 
fundamental building blocks) are for me the most 
fundamental building blocks of reality, at least that 
we know of in the present-day situation.

\section{The constancy of the speed of light}
\label{sec:constspli}

Observations confirm that the speed of light in vacuum is 
approximately 300 000 000 m/s, in agreement with what 
Maxwell's equations predict. But if Maxwell's equations 
hold for all observers in constant uniform motion, does 
this not mean that \textit{same} light beam moves at the speed of 
300 000 000 m/s relative to \textit{each} observer?\\

If it were so, this would not be consistent with how we 
normally perceive our everyday world. Because if I am on 
a train and throw a stone in the direction that the train 
travels, then the stone does not move relative to the 
ground with the same speed as the stone moves relative to 
the train. Relative to the ground the stone instead moves 
with the speed of the stone relative to the train \textit{plus} the 
speed of the train relative to the ground, i.e., with a 
speed that is \textit{greater} than the speed of the stone relative 
to the train. Suppose further that the horn on the train 
emits a soundwave in the direction that the train travels. 
Since the sound medium, i.e., the air, is at rest relative 
to the ground, the sound wave travels relative to the ground 
with the speed of sound (about 340 m/s), and that regardless 
of the speed of the train. But relative to the train, the 
sound wave travels with the speed of sound \textit{minus} the speed 
of the train, i.e., with a speed that is \textit{lower} than the 
speed of sound.\\ 

That the same light beam would travel with the same speed 
regardless of which observer measures the light beam's 
speed, is remarkable if one considers what classical 
physics predicts, or simply what our everyday experience 
tells us. So how can one explain that light moves with 
the same speed relative to all inertial reference systems?\\

But what does one really mean when one says that light
moves with the "same speed" relative to all inertial 
reference systems? For one cannot measure the speed of 
light by sending a light signal between two clocks located 
in two different places in space and then divide the 
distance with the time difference on the clocks, if one does 
not also have \textit{synchronized} clocks. To calculate the time 
difference, the clocks need to show the \textit{same time at the 
same time}, i.e., \textit{simultaneously}. This means that the 
clocks must be \textit{simultaneous} or, in another word, 
\textit{synchronized}.\\

Simultaneity and synchronization of clocks are important 
and key concepts for the understanding of the special 
theory of relativity. We therefore need to be careful what 
we mean by these concepts. One needs to do synchronization 
in a systematic and reliable way. Einstein chose to 
synchronize clocks by sending light beams between the 
clocks. Since he had \textit{postulated} that the speed of light is 
constant, he could use this to synchronize clocks. (It seems 
perhaps more natural, simpler and more intuitive to simply 
synchronize clocks at the same place in space and then 
distribute them out to the places in space where one wants 
them to be. But if one starts to move around clocks in this 
way, one will have practical problems, e.g., with the fact 
that time passes more slowly for clocks in motion, which we 
know is the case on the basis of experiments and
observations.)\\

\begin{figure}
\includegraphics[scale=.8]{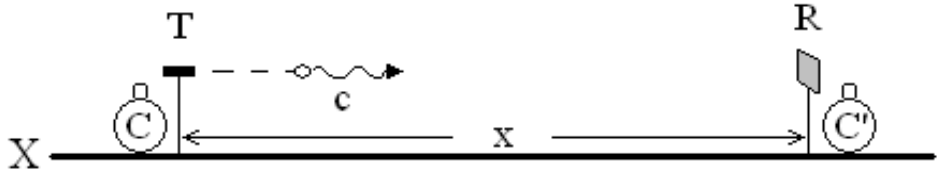}\\
\caption{Determining the speed of light.}
\end{figure}

The \textit{speed of light} one 
determines by sending a light beam from 
a transmitter $T$ to a reflector $R$, e.g., a mirror, which 
reflects the light back to $T$ (see figure~4). Let $x$
be the distance between $T$ and $R$, both of which are at rest 
relative to a single reference body $X$. The distance $x$ is 
measured by placing measuring rods on $X$ from $T$ to $R$. The 
time interval $t$ between that the light was send out and returned 
is then measured by the \textit{same} clock $C$, 
which is at rest next 
to $T$ on the reference body $X$. The distance back and forth, 
i.e., twice the distance $2x$ between $T$ and $R$, is then
divided by the time interval $t$ measured on $C$. The speed 
$c = 2x / t$ one then obtains, is \textit{defined as the speed of 
light}. Defined in this way, the speed of light can be 
measured and is thus a meaningful concept in the spirit of 
Mach and Einstein. It is this speed that Einstein postulated 
to be constant.\\ 

Note that one cannot \textit{prove} that a postulate 
is true. One can only convince oneself of its correctness by 
testing the postulate in more and more different situations. 
And so far the postulate has never turned out to be wrong, 
so in that sense it is a satisfactory postulate.\\

Once one has defined and measured the speed of light $c$
with the above described method, one can then use it to 
\textit{synchronize clocks}. Assume that one has two clocks $C$ and 
$C'$ that one wants to synchronize relative to a reference 
body $X$. Both clocks must then be at rest relative to $X$. 
The clocks are at a distance $x$ apart on $X$. At a given 
time $t$ on $C$, say $t = 0$, a light beam is sent from clock 
$C$ to $C'$. When the light beam reaches $C'$, one sets $C'$ to 
show the time $t' = x / c$. In this way one can synchronize 
any clocks at rest on a single reference body (inertial 
reference system).\\

What does it mean then that two events are \textit{simultaneous}? 
Simultaneity is \textit{defined} as follows: Assume that we have 
clocks that are synchronized relative to an inertial 
reference system $X$. Further assume that two events, such 
as two lightning strikes, occur at two different places 
in space. If the synchronized clocks of $X$ at these two 
places in space show the same time for the events, one 
says that the events are \textit{simultaneous} relative to the 
inertial reference system $X$.\\

But how does one know that the light beam \textit{really} 
arrives at the reflector after "half the time" when 
one synchronizes two clocks with the above described 
method of synchronization? For could the light speed not 
be higher in one direction and lower in the other?\\ 

That depends a little bit on what one means. Assume 
that one on a reference body $Y$ \textit{already has synchronized 
clocks} with the above given method. Assume further that 
on a reference body $X$ one intends to synchronize two 
clocks, where $X$ moves relative to $Y$ with 
velocity $v$ (see figure~5). The 
synchronization situation in $X$ is observed from the 
reference body $Y$. Relative to $Y$ the reflector $R$ moves 
either towards or away from the light beam, depending 
on how $X$ is moving relative to $Y$. According to the 
clocks on $Y$ the light beam will not arrive at the 
reflector when the clock $C$ shows half the time, i.e., 
$t / 2 = x / c$. This means that observers on $Y$ do not 
think that $C$ and $C'$ on $X$ have become synchronized. But 
according to the clocks $C$ and $C'$ on $X$ the light beam 
will, \textit{by definition}, arrive at $R$ after half the time. We 
have thus demonstrated that with the above definition of 
simultaneity, it becomes a \textit{relative} concept. If two 
events are simultaneous relative to one inertial 
reference system, these two events in general will not 
be simultaneous relative to other inertial reference 
systems in motion relative to the first inertial 
reference system.\\

\begin{figure}
\includegraphics[scale=.72]{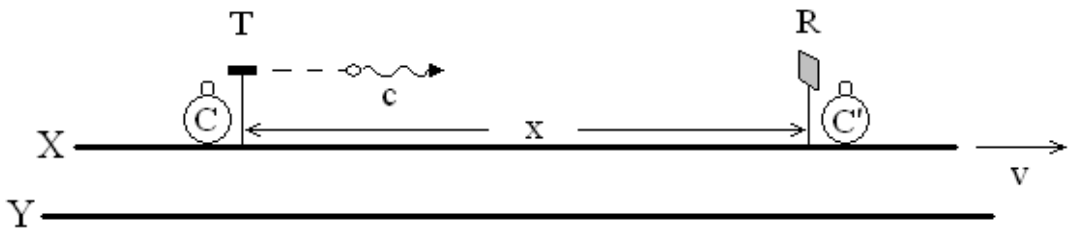}\\
\caption{The relativity of simultaneity.}
\end{figure}

But now I do not think this is what one really thinks of 
and means with the above question, but whether the light 
beam \textit{really} hits the reflector after "half 
the time." One asks how it \textit{really} or 
\textit{truly} is, i.e., if the light beam 
arrives at the reflector after half the time \textit{objectively 
seen}?\\ 

However, this is a question that we strictly seen 
cannot answer. The only thing we can do is to make use of 
synchronized clocks to check whether this is the case or 
not. Because, we use the definition of the speed of light 
to synchronize clocks. This definition involves a 
time-interval on one and the same clock $C$ located at $T$. 
In other words, the definition of the speed of light 
does not involve the clock $C'$ at $R$. \textit{First} we define and 
determine the speed of light using clock $C$. \textit{Then} the 
speed of light is used as a tool to synchronize clock $C'$. 
In that moment when the light beam is reflected on $R$, the 
clock $C'$ is set to show the time $t' = x / c$. But 
$t' = x / c$ is half of the time $t = 2x / c$, where $t$ is the 
time interval between the emission and the return of the 
light beam at $T$, and $t$ was measured on $C$. Clock $C'$ is thus 
set to show half the time, i.e., $t / 2$, when the light
beam is reflected on $R$. The light beam therefore hits the 
reflector after half the time by definition.

The only thing we can do is to define the speed of light 
and then synchronize clocks with the above described 
method. Then there are no guarantees that this systematic 
and consistent way of synchronizing clocks will work in 
practice. That is up to experimental tests to settle. But 
so far, this way of synchronizing clocks, and to define 
space and time, has proven to work excellent in practice. 
If this had not been the case, then one would have had to 
try something else. 

Then it may be the case that one 
perhaps could in different ways motivate why it would be 
highly unlikely that it should not work, or that there would 
have been severe consequences for one's concepts, theories and
understanding of reality if it had not worked, and that 
one therefore expects that it should work. But we are 
not in the position to tell reality how it should work. 
Our job is, on the basis of empirical tests and 
investigations, to find out how reality works.\\

From the \textit{postulate} on the constancy of the speed of 
light and the \textit{postulate} that all laws of nature are the 
same in all inertial reference systems, one can now 
derive the special theory of relativity. From only these 
two postulates follows \textit{exactly the same} equations that 
Lorentz arrived at, i.e., the Lorentz equations, but with 
a different conceptual and philosophical basis.\\
 
From the Lorentz equations it follows, among other 
things, that time goes slower, and that bodies become 
shorter in the direction of motion, for bodies in motion 
relative to other bodies. But also that simultaneity 
becomes a relative concept. In addition to these 
relativistic effects, there are several other things that 
follow from the special theory of relativity, e.g., that 
quantities such as momentum and energy are changed. The 
famous equation $E = mc^2$ was presented by Einstein in the 
wake of the special theory of relativity. However, we will 
here mainly focus on how the notions of space, time and 
simultaneity came to change with the special theory of 
relativity.

\section{A Lorentzian interpretation of the special theory of relativity}
\label{sec:lorintspecrel}

How should one then \textit{explain} and 
\textit{understand} the above 
mentioned relativistic effects? That 
depends on how one \textit{interprets} the special theory of 
relativity. Let us first do as Lorentz did and assume that
for all observers that are in motion relative to Newton's absolute 
space, or the ether, time and length change according to
the Lorentz equations. The faster one moves relative to the 
absolute space, or the ether, the slower time goes and lengths 
become shorter in the direction of motion. The combined 
effect of these two effects, is that the speed of light 
becomes the same relative to all inertial reference systems.\\

But if time is goes slower and distances become shorter, as 
seen from one inertial reference system at rest relative to 
Newton's absolute space, or the ether, does that not mean 
that time goes faster and that distances become longer when 
seen from all other inertial reference systems in motion 
relative to Newton's absolute space, or the ether? And if 
this is the case, would one then not be able to determine 
who actually is in motion and who is not, in violation with 
one of the postulates of the special theory of relativity?\\

Well, if this was the case, then one would have a means of 
detecting absolute motion. But in fact the situation will be 
completely \textit{symmetrical} for all inertial reference systems. 
What enables this, apart from the fact that time and length 
change, is that also \textit{simultaneity} has become a relative 
concept. Also in the "Lorentzian interpretation", we use 
the above definition of simultaneity. All inertial 
reference systems, regardless whether they are in absolute 
rest or not, will find that their clocks are synchronized 
and therefore simultaneous. But as we have already seen, if 
two inertial reference systems are in motion relative to one 
another, it generally holds that two events that are 
simultaneous relative to one of the inertial reference 
systems, are not simultaneous relative to the other inertial 
reference system. No observers on these inertial reference 
systems can therefore distinguish between constant uniform 
motion relative to the absolute space, from being at rest 
relative to the absolute space. From the observers' point
of views, their situations are completely symmetrical 
or equivalent.\\

But according to the "Lorentzian interpretation", or 
"Lorentzian approach", it is only \textit{seemingly} 
so. Because, as seen from "the point of view of reality", 
or \textit{objectively seen}, their situations are in fact 
asymmetrical. This asymmetry has its origin in the fact 
that it matters who objectively seen moves or does not 
move relative to the absolute Newtonian space, or the 
ether. For the observer who happens to be in absolute 
rest, two events that are simultaneous for this observer 
are also \textit{absolutely simultaneous}. I.e., only for observers 
in absolute rest, will events that they \textit{observe} to be 
simultaneous, also be simultaneous \textit{objectively seen}. For 
all other observers in motion relative to the absolute 
space, the events that they observe to be simultaneous, 
will objectively seen only seemingly be simultaneous. 
Because, having adopted a "Lorentzian interpretation" of 
the special theory of relativity, one has hence also 
assumed \textit{absolute simultaneity}. 

But that is not to say that 
the observers who live in a reality where absolutely 
simultaneity prevails, can distinguish, based on the 
definition of simultaneity which they themselves have 
created, between events which are absolutely simultaneous 
and events which objectively seen only are seemingly 
simultaneous. The light beam used in the synchronization 
method described above, will \textit{objectively seen} reach the 
reflector after half the absolute time it takes for the 
light beam to go from and come back to the transmitter, 
only if the transmitter and reflector are in absolute 
rest (relative to the absolute 
Newtonian space). But even with a Lorentzian interpretation, 
this is not something that can be measured by an observer, 
since also in the Lorentzian interpretation all inertial 
reference systems are equivalent; i.e., there is no way 
for an observer to find out who really is in absolute rest 
or who really is in a state of absolute motion.\\

With a Lorentzian interpretation one can thus \textit{explain} and 
\textit{understand} how the speed of light, but also how all other 
laws of nature, are, or appear to be, the same relative to 
all inertial reference systems. But to do so, one needs to 
"go outside reality" and "see it from the outside", i.e., 
from a meta perspective. However, this is not a
perspective that is accessible to observers, who are 
(by definition) limited to observe reality "from within". 
The Lorentzian explanation of the constancy of the speed of 
light, therefore includes assumptions and concepts which are 
not based on observations. One is forced to introduce an 
abstract idea of an absolute Newtonian space which cannot
be observed. Events are only truly simultaneous in
relation to this absolute space. For all observers that 
move relative to this absolute space, time goes \textit{objectively 
seen} more slowly and distances become \textit{objectively seen} 
shorter in the direction of motion.\\

At first sight, the Lorentzian interpretation may seem 
radical. But it is a natural interpretation to do, when 
trying to adapt our everyday experiences and the classical 
theories, to the observational facts that the speed of
light and the natural laws are the same for 
all inertial reference systems, if one at the same time 
wants to maintain a Newtonian approach to reality. Instead 
of a "Newtonian approach to reality", one could also say
an "everyday-experience-based world view." That would make 
it more clear that the Lorentzian interpretation is
closely and intuitively connected with everyday concepts 
and the possibility to translate the predictions of the 
special theory of relativity to something more intuitive 
and understandable for us human beings. Because the 
Newtonian and Lorentzian conception of the world enable 
us to create and imagine a more understandable and 
intuitive internal mental picture or model of reality.

\section{The Lorentzian interpretation - A train example}
\label{sec:atrainex}

To make what have been said above more concrete, we will 
now, by considering an example, look more closely at how 
a Lorentzian interpretation provides an \textit{explanation} of 
how the speed of light and the laws of nature can be the 
same for all observers in constant uniform motion. For 
the sake of simplicity we restrict ourselves to consider 
only one space dimension $x$. We lose nothing in generality 
by doing so, because we can always turn the three space 
coordinate axes $x$, $y$, $z$, so that the motion is only in 
the $x$-direction.\\

On board one of the cars on a train, is a person at the 
rest in that car's rear end, in relation to the direction 
that the train moves. This person sends a beam of light 
towards the front end of the car, where it is reflected 
back by a mirror. The person measures with a clock $C$ the 
time interval from the point in time when the light beam 
is sent out until it comes back again. The length of the 
car x can be measured by the person on board by placing 
measuring rods along the floor of the car.\\

We assume now that the embankment is in absolute rest 
relative to the absolute space. Relative to the 
embankment the light beam objectively seen thus moves 
with the speed of light, i.e., $c =$ 300 000 000 m/s. 
Assume that the train is travelling at a speed $v$ which 
is close to the speed of light. Seen from the embankment, 
the mirror is moving away from the light beam (see 
figure~6).\\

\begin{figure}
\includegraphics[scale=.8]{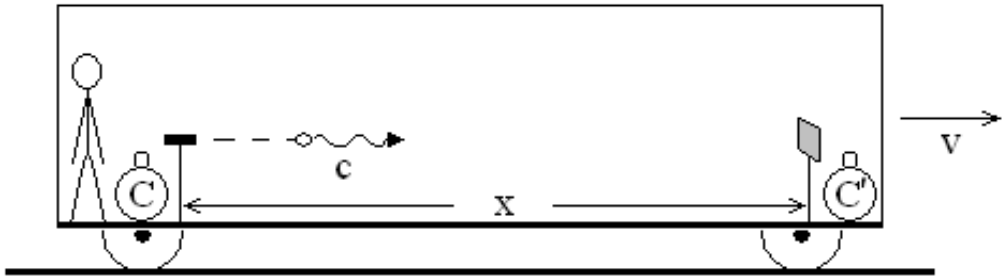}\\
\caption{Train-embankment inertial reference systems.}
\end{figure}

If nothing would happen with the lengths or with the time 
on the clocks on board the train, then a simple calculation 
shows that it would take a longer time for light to travel 
back and forth along the car of the train, \textit{as seen both 
from the embankment as well as from the person on board 
the train}. (This even though the person and the clock $C$ on 
board the train move, as seen from the embankment, towards 
the light beam after it has been reflected on the mirror.) 
This means that the person on the train would have to wait 
a longer time for the light beam to come back. Since we
have assumed that lengths on the train do not change, this 
would mean that the person on the train would have measured 
a lower value on the speed of light than 300 000 000 m/s. 
This would contradict the postulate that the speed of light 
is to be the same relative to all inertial reference 
systems, as we know from experiments that it must be.\\
 
This means that lengths and/or time must change in some 
way to compensate for the longer time the light beam takes 
to travel back and forth along the car of the train. If 
lengths become shorter and the time on the clocks on board 
the train moves slower, relative to the embankment,
precisely in the way that the Lorentz equations prescribe, 
then one can by a simple calculation show that this exactly 
compensates for the longer time that the light beam takes 
to go back and forth along the car of the train. The result 
or combined effect of this, is that the person on board will 
measure the speed of light to be 300 000 000 m/s after all.\\
 
The faster the train moves, the shorter lengths on the train 
will become, and time will go slower and slower, relative to 
the embankment. And if the person on the train instead would 
be in the front end of car and sends the beam to the rear 
end of the car, where it is reflected, a similar reasoning 
would give the same result or conclusion, i.e., the speed of 
light remains invariant.\\

But if time goes slower and distances become shorter on the 
train relative to the embankment, does that not mean for the 
person on the train, that time goes faster and distances 
become longer on the embankment? If this was the case,
that would contradict the postulate which says that reality 
must look the same from all inertial reference systems. This 
means that the situation as seen from the train, must be 
identical to the situation as seen from the embankment, apart 
from the fact that the embankment moves in the opposite 
direction. (The size of the relative speed must however be 
the same, seen from both of the reference bodies. For if 
this were not the case, then the two reference bodies would 
not be equivalent, which again would violate one of the 
postulates of the special theory of relativity.)\\ 

Therefore, seen from the train, time on the embankment must go slower 
than on the train, and lengths on the embankment must become 
shorter in the direction of motion. But this is also possible, 
thanks to the fact that \textit{simultaneity} has become a relative 
concept. For in the case with the train described above, 
there is an asymmetry between the situation on the train and 
the situation on the embankment. The asymmetry has to do 
with how times and distances are measured. As have already 
been said, the person on the train measures lengths on the 
train by placing measuring rods along the floor of the car 
of the train. Furthermore, this person measures time on his 
clock in the rear end of the car. Therefore, as seen from 
the train, the point in time of the emission of the light 
beam, as well as the point in time of its return, are both 
measured at the \textit{same place} in space.\\

The measuring situation \textit{seen from the embankment} is however 
different. To measure, e.g., the length of the car of the 
train from the embankment, one cannot place measuring rods 
along the ground from one end of the car to the other. The 
reason for that is that the train is moving and therefore 
would have moved while one did this. One would then not 
have measured the length of the car, but simply something 
else. Nor is the time between the emission and return of 
the light beam possible to measure from the same place on 
the embankment. Because the event when the person on the 
train emits the light beam and the event when the beam 
comes back to the person on the train, occur at \textit{two 
different places} along the embankment.\\ 

Therefore one must 
use two different clocks for these two places on the 
embankment. It is here that the concept of simultaneity 
comes in, since the clocks must be synchronized. And 
synchronization and simultaneity are basically the same 
thing. According to what have been said above, the observers 
on the two reference bodies will not agree on 
which events are simultaenous and which are not. Since 
they move relative to one another, two events that are 
simultaneous relative to the train, will \textit{not} be 
simultaneous relative to the embankment; and vice versa. 
The same thing holds for synchronized clocks. Two clocks 
that are synchronized relative to the train, will \textit{not} be 
synchronized relative to the embankment; and vice versa. 
And this again because they move relative to one another.\\

Let us see how the synchronization situation becomes in 
the case with the train. Again we assume that a clock $C'$ 
is located by the mirror. When the light beam is reflected 
on the mirror, then $C'$ is set to show the time $t' = x / c$, 
where we have assumed that the light beam was sent at the 
time $t = 0$ on clock $C$. The time $x / c$ is, according to 
clock $C$, precisely \textit{half the time} it takes for the light 
beam to go back and forth along the car of the train.\\ 

But as seen or measured on clocks that are at rest on the 
embankment, the light beam is not reflected after half the 
time between it was sent and returned. According to times 
measured by clocks at rest on the embankment, the reflection
 occurs after a time which is \textit{longer} than half of the time 
interval between the emission and the return. The reason 
for this is that clock $C'$, as seen from the embankment, is 
moving \textit{away from} the light beam \textit{before} 
the reflection, while the clock $C$ is 
moving \textit{towards} the light beam \textit{after} 
the reflection. According to the embankment, it therefore 
takes a longer time for the light beam to reach $C'$ from $C$, 
than it takes for the light beam to return back to the $C$ 
from $C'$. Relative to the embankment clock $C'$ is hence not 
synchronized at the right time. Because relative to the 
embankment, the clock $C$ does not show half the time, i.e., 
$x / c$, at the same time that the light beam is reflected. 
According to the embankment, the light beam has not yet 
reached the mirror when clock $C$ shows the time $x / c$. As 
seen from the embankment, once the light beam reaches the 
mirror and clock $C'$ is set to show the time $x / c$, then 
clock $C$ shows a time which is later than $x / c$. As seen 
from the embankment, the clocks on board are thus not synchronized.\\

But observers on board the train insist that the clocks 
$C$ and $C'$ are synchronized. The observers on board the 
train instead consider the clocks on the embankment as not 
synchronized. Hence observers on the train and observers on 
the embankment, do not agree on whose clocks are synchronized 
and therefore not which events that are simultaenous. Both 
inertial reference systems consider their clocks to be 
synchronized, which they also are completely entitled to think. 
The thing is that there is no way for observers on board the 
train or the embankment to determine which 
of them \textit{really} are 
right. Thus they cannot through measurements determine who 
\textit{really} are in absolute rest. Now we happen to know, since 
this was assumed above, that it is the embankment which really 
is in absolute rest. But that is only because we view the 
situation from an \textit{objective perspective} 
which is not available 
for the observers on the train or the observers on the 
embankment; or for any observer at all for that
matter. This meta perspective is possible thanks to that
we have assumed that the "Lorentzian interpretation" or 
"Lorentzian world view" is true. This interpretation thus 
offers an \textit{explanation} 
for how reality manages the feat to get 
all observers in constant uniform motion to agree that the 
speed of light is always the same.\\

If one on the embankment wants to measure, e.g., the length of 
the car of the train while it is in motion, one uses clocks 
that are \textit{synchronized} relative to the embankment's inertial 
reference system. The length of the car is measured by 
measuring the position of car's rear and front end, according 
to these synchronized clocks, \textit{simultaenously 
in two different places along the embankment}.\\

In a similar way one determines, in the embankment's inertial 
reference system, the time between the event when the
light beam leaves $C$ and the event when it returns to $C$. First 
one identifies the two places along the embankment where the 
light beam left and came back to clock $C$. The points in time of 
these two events, one reads from clocks that are synchronized 
relative to the embankment. The time between these two events 
is then obtained by simply taking the difference of these two 
points in time. Seen from the embankment, it is true that 
these two points in time are not measured simultaenously, 
but with clocks that are synchronized relative to 
the embankment.\\

However, observers on board the train do not consider the 
clocks along the embankment to be synchronized relative to 
one another. According to the train's inertial reference 
system, the clocks on the embankment have therefore measured 
where the front and rear of the car of the train 
are \textit{at two different points in time}. So 
observers on the train do not 
consider it to be the length of car that the observers on 
the embankment have measured up, but something else.\\

According to the "Lorentzian interpretation" of the Lorentz 
equations, it is thus only for those observers who are in 
absolute rest that the speed of light \textit{really} 
or \textit{objectively seen} is 300 000 000 m/s, 
regardless of the speed of the light 
source and regardless of the direction in which the light 
travels. For all other observers, who are in motion relative 
to the absolute space, it is only \textit{seemingly} so. This 
"illusion" is thus made possible by the fact that 
time \textit{really} goes slower 
and that lengths \textit{really} become shorter, relative 
to the time and lengths of the absolute space, for reference 
bodies that \textit{really} move relative to the absolute space. The 
relativity of the concept of simultaneity enables observers on 
board the train to perceive the situation as if they are
at rest and that it is the embankment which moves relative 
to them (with the same relative speed as seen from the 
embankment). Space and time will thus change in the same 
way for all observers who are in constant uniform motion 
relative to each other. One can demonstrate this through 
the systematic use of the Lorentz equations and by
applying the definitions of space, time, and simultaneity, 
on the one hand relative to the embankment, and on the 
other hand relative to the train.\\

Does then the Lorentzian interpretation give any 
explanation as to \textit{why} time goes more slowly 
and \textit{why} 
lengths become shorter for observers who are in 
absolute motion? No, the Lorentzian interpretation 
simply says that this is how reality works. On the 
other hand, if we assume that this is how reality 
works, then it would not be very difficult to come 
up with all kinds of reasons and explanations to why 
it could be like that. Since we have already assumed 
an existence of an absolute space which we cannot 
observe, why should we then also not be able to give 
this absolute space characteristics or properties 
that would make bodies, moving through this space, 
shorter and time on them go slower? But it does not 
necessarily have to be properties of the absolute 
space. One could imagine all sorts of properties of 
reality, which would be the reasons why time on clocks 
goes slower and why distances become shorter for bodies 
in absolute motion. It is, of course, not impossible 
that such characteristics of reality one day might be 
observed. On the other hand, since we already have 
assumed the existence of an abstract absolute space 
that we cannot observe, it is not even sure that we 
feel, or consider it to be necessary, that we should 
actually be able to observe everything which according 
to our models and theories are assumed to exist in reality.\\
 
Once again we here touch upon questions having to with the 
philosophy of science. However, let me just conclude by 
saying that, despite its abstract and non-observable 
nature, I do not feel or consider the concept, and 
the assumption of the existence, of an absolute space to 
be something far-fetched or arbitrary. Instead I consider 
it, together with the concept of absolute simultaneity, 
in many respects to be something natural, and intuitive. 
Because remember that we had perhaps not even questioned 
the notion of absolute space, time, and simultaneity, if 
it were not for philosophers and physicists such as Mach 
and Einstein. For it was thanks to that Mach questioned 
and criticized the concept or idea of absolute 
space, and that Einstein was working on the basis of, 
and applied, this philosophical approach to problems 
directly and indirectly connected to the nature of 
light, which enabled the development of the theory of 
relativity, and much of modern physics as we know 
it today.

\section{Operational special theory of relativity}
\label{sec:opspecrel}

But now it was not a "Lorentzian interpretation" that 
Einstein did of the special theory of relativity. Einstein 
was critical of, what we here have called, the "Lorentzian 
interpretation". It was not length contraction and time 
dilation in itself that Einstein considered to be 
unsatisfactory, for these phenomena are also found in 
the special theory of relativity. It was the 
\textit{interpretation} or \textit{explanation} 
of these phenomena that 
Einstein found unsatisfactory. He thought that Lorentz's 
explanation, that lengths become shorter and time goes 
slower as a result of their movement relative to the 
absolute space, or ether, was unsatisfactory. In the 
Lorentzian interpretation the length contraction and 
time dilation are absolute in a Newtonian sense, while 
Einstein had a different approach on the whole thing with 
the special theory of relativity. Einstein wanted to get 
away from Newton's absolute space, the ether hypothesis, 
and more generally concepts and hypotheses that do not 
have its basis in observations. He realized that it was 
the notions of space, time, and simultaneity, that needed 
to be changed, if one wanted to get away from the 
assumption of an existence of a non-observable and 
abstract concept such as the absolute Newtonian space, 
or the ether.\\
	
Einstein attacked the problems having to do with space, time, 
simultaneity, the invariancy of the speed of light 
and the other laws of nature, in an \textit{operational} 
way. In its original form the special theory of relativity 
is an \textit{operational theory}, which is important to remember 
when trying to understand the special theory of relativity 
and its predictions. By \textit{operational} I mean, that one does 
not explain natural phenomena in terms of non-observable 
and abstract concepts, in a way similar to how one 
explains the pressure of a gas by assuming the existence 
of atoms, or the wave nature of light by assuming the 
existence of an ether. Instead one starts from 
observations of reality and describes how other natural 
phenomena can be described using these observations. All 
concepts in the theory are defined on the basis of 
something which in a concrete way can be observed and 
measured, and predictions from the theory only have a 
meaning if they can be observed and measured. The theory 
does not go, so to speak, beyond experience.\\ 

In its 
original form, the special theory of relativity is, e.g., 
not dependent on the concept of \textit{spacetime}. The theory is 
free from interpretation and its predictions are very 
concrete. One could describe the special theory of 
relativity as taking a step back from the Lorentzian 
interpretation, in the sense that it assumes less and 
is therefore more general than the Lorentzian 
interpretation. The Lorentzian interpretation can be seen 
as \textit{one} possible interpretation of the special theory of 
relativity. This is not to say that the special theory of 
relativity needs an interpretation, because strictly seen 
it does not.\\

So Einstein did not really try to \textit{interpret}, 
or to \textit{explain how} or \textit{why} 
the speed of light and the laws of nature are 
invariant with respect to all inertial reference systems. 
He simply \textit{postulated} that it was so, i.e., he took 
it as an observational fact and assumed that it was true. The 
concepts of space, time, and simultaneity, he \textit{defined} 
what they should be. Distances in space are measured with 
measuring rods, relative to a reference body at rest, i.e., 
an inertial reference system. Time is what one measures 
with ordinary clocks. When measuring time in the special 
theory of relativity, in general one does not move the 
same clock around in space. Instead one places a clock 
at every position in space, which in principle is possible 
to do. The clocks are assumed to be of identical 
construction. Furthermore, one assumes that time goes at 
the same rate (i.e., equally fast) on every clock in space.\\

In the special theory of relativity, are thus space, time, 
and simultaneity, operationally defined concepts. Strictly 
speaking the theory offers no explanations of \textit{how} reality 
objectively seen (ontologically) actually works, or \textit{why} it 
works as it does. The only thing the theory says is that, 
if we define space, time, simultaneity, and the speed of 
light in the above described way, and assume that the speed 
of light and all laws of nature are invariant with respect 
to all inertial reference systems, then it follows that 
one will measure and observe that space and time in 
different inertial reference systems are related in the 
way that the Lorentz equations prescribe. Why and how 
reality manages the feat to make the speed of light and 
the laws of nature to be invariant with respect to every 
inertial reference system, are questions that the special 
theory of relativity strictly speaking does not answer. As 
have already been said, it assumes or postulates that this 
is the case. And these two postulates have been verified
in countless experiments and observations.\\

It is true that the special theory of relativity strictly 
speaking requires no interpretation. However, personally 
I think that it is difficult to stop at an operational 
account of the special theory of relativity, without the 
brain wanting to jump to conclusions, try to create an 
overall picture of reality, or start to think about 
underlying explanations, ontologically seen, to how 
reality works, but also why it works as it does. How it 
can be that the speed of light speed, and the laws of 
nature, are the same for all inertial reference systems. 
And how reality manages this feat.\\ 

In its original 
operational form, the special theory of relativity is to 
me unsatisfactory in a similar way that I find 
thermodynamics and quantum mechanics to be unsatisfactory. 
Neither thermodynamics or quantum mechanics give an
ontological picture or explanation, of what reality looks 
like or how it behaves, on the microscopic level. Both 
these theories are more operational to their nature. 
(The so-called "Copenhagen interpretation" of quantum 
mechanics is in my opinion not really an interpretation, 
but it rather concerns what quantum mechanics
operationally has to say.) However, unlike thermodynamics 
and quantum mechanics, a theory such as statistical 
mechanics, combined with classical mechanics and the 
atomic hypothesis, in principle gives a complete 
description of reality. In this respect, the special 
theory of relativity resembles the operational theories 
of thermodynamics and quantum mechanics more, than it 
does a theory such as statistical mechanics.\\
	
An interesting comparison can here be made with 
what Einstein himself thought of quantum mechanics, i.e., 
as unsatisfactory because it did not provide a complete 
ontological description of reality. Bohr is said to have 
been inspired by the operational character of special theory 
of relativity when he participated in and contributed to 
the development of quantum mechanics. He therefore thought 
that Einstein would like quantum mechanics, since also this 
was an operational theory. But it seems as if Einstein in 
the years between the creation of the special theory of 
relativity and the creation of quantum mechanics, had partly 
changed attitude, or philosophy. From being more operational, 
in the spirit of Mach, to becoming more accepting of more 
not directly observable elements and concepts in the theories, 
such as the spacetime concept.\\ 

But probably had there in 
Einstein always been certain notions and opinions that he 
was not prepared to alter or give up on. Philosophically 
seen, Einstein seems to fundamentally have had an ontological 
approach to reality. He was, e.g., not willing to give up 
on an objective description and understanding of reality 
when it came to quantum mechanics. Maybe one also could take 
the finiteness of the universe as an example of something 
that Einstein for a while had a difficult time to accept, 
despite the fact that his own general theory of relativity 
rendered such a universe possible. Probably because it was 
contrary to his inner beliefs and convictions of how reality 
reasonably should be.\\

If one wants to, one can thus refrain from trying to 
interpret the special theory of relativity. This is in some 
sense similar to how many physicists in the present-day 
situation approach quantum mechanics. But for me this 
approach is unsatisfactory. I find 
quantum mechanics as unsatisfactory as a final description 
and explanation of reality. Although I personally do not 
believe so, at the moment quantum mechanics may be adequate, 
and perhaps even inevitable. But it does not agree well 
with what I think should be the ultimate ambition and 
final goal of science, and that is to understand how 
reality objectively or ontologically seen is and works. 
Then it is possible that this ambition and goal never 
will, or cannot be accomplished. But science should not 
give up on this ambition until there are very good and 
convincing reasons for doing this. And any such reasons 
I cannot remotely see in the present-day situation.

\section{The spacetime interpretation}
\label{sec:sptint}

One way to look at the concept of \textit{spacetime}, is as a 
practical way to illustrate what happens in space and time. 
E.g., at train traffic supervision centers one uses a 
kind of \textit{spacetime diagram} to facilitate the monitoring of 
the trains. Such a diagram has two perpendicular axes, where 
space is on one axis and time on the other axis. A train's 
movement is represented by a line, where the slope of the 
line gives the speed of the train. An intersection of two 
lines corresponds to a meeting between two trains. If the 
trains did not meet each other on two separate tracks, it 
means that a collision between the trains has occurred. A 
spacetime diagram of this kind is, of course, nothing 
mysterious or fundamentally new, but it is just a 
practical way to illustrate the movement of trains (see 
figure~7).\\

\begin{figure}
\includegraphics[scale=.8]{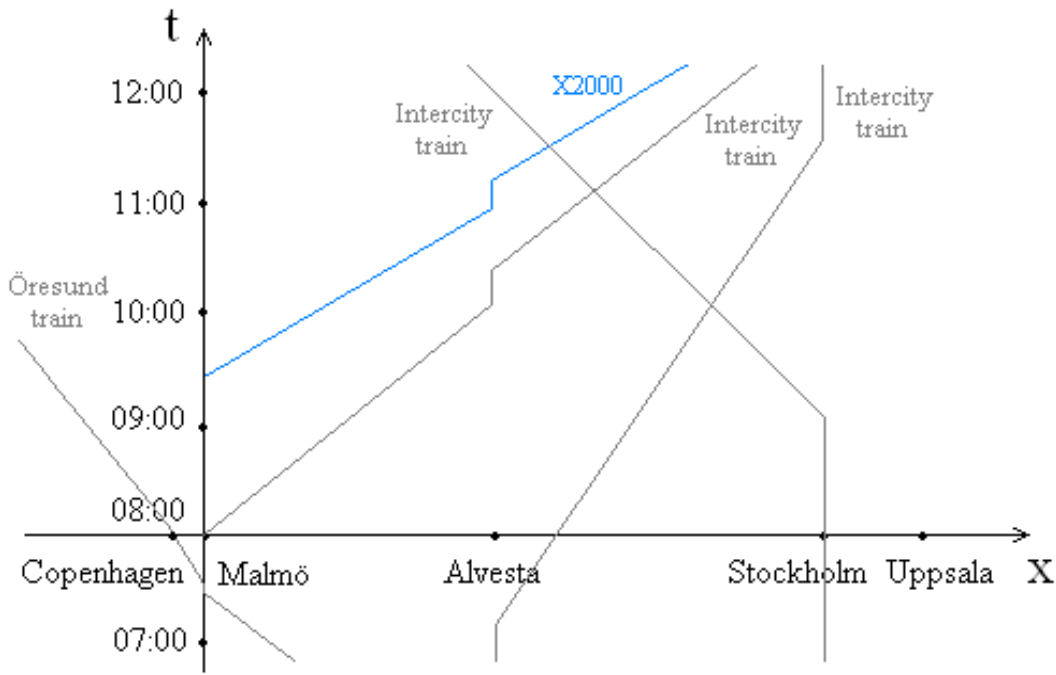}\\
\caption{An example of a "spacetime diagram" at a train 
traffic supervision center.}
\end{figure}	 

Instead of trains, one could just as well illustrate the 
motion of atoms in this way. The motion of atoms would then 
be shown as lines in the diagram, and if two lines intersect 
it means that a collision between the atoms has occurred. If 
the atoms can move in two space dimensions, the spacetime 
diagram would be three-dimensional, i.e., two space axes and 
one time axis. And if the atoms can move in three space 
dimensions, the spacetime diagram instead becomes 
four-dimensional. Although it in practice is a greater 
challenge to illustrate a four-dimensional space in a 
pedagogical way, there is nothing strange with a 
four-dimensional space. We all live in a four-dimensional 
world, i.e., with three space dimensions and one time dimension. 

If I say that I am at the main entrance to the 
Turning torso in Malm\"{o}, at 12:00 on the 1st of January 2009, 
I have in principle described where I am with four coordinates, 
i.e., three space coordinates (giving the location in space) 
and one time coordinate (the point in time when it happens). 
If one would place the space origin at the main entrance
to Turning torso, then my spacetime coordinates would be 
$(x, y, z, t) = (0, 0, 0, 12)$. 

It takes four coordinates to 
describe the motion of an object in space. One could, e.g., 
describe the motion of a 200-meter runner when he or she 
runs on the finishing stretch as $(x, y, z, t) = 
(v \cdot t, 0, 0, t)$, where $v$ is the runner's speed and $t$ is 
the time the runner has been on the finishing stretch. 
Since the motion on the finishing stretch in practice 
takes place along a straight line in space, one can 
illustrate the runner's motion as a line in a 
two-dimensional spacetime, with one space axis and one 
time axis. With this two-dimensional spacetime the 
coordinates would then be $(v \cdot t, t)$. Again, the spacetime 
concept used in this way, is not or does not involve 
anything mysterious or fundamentally new. It is just a 
convenient and alternative way to illustrate the 
motion of bodies.\\

When it comes to the spacetime concept in the special 
theory of relativity, there is nothing that prevents 
one from limit oneself to consider and use spacetime 
in the above described manner, i.e., as a practical 
and useful way to illustrate what happens in space 
and time. It seems to me that many physicists, to a 
large extent, perceive and understand the spacetime 
concept only in this way. I think that has to do with 
the fact that they either are not aware of any other 
way to understand the concept of spacetime, or 
because they simply do not have a clear understanding 
of the meaning of spacetime concept that we will 
consider below.\\ 

Admittedly, the special theory of 
relativity requires no other interpretation of the 
spacetime concept other than the one described above. 
But there is another interpretational possibility of 
the concept of spacetime that is natural to adopt, if 
the special theory of relativity is to go from a purely 
operational theory, to a complete and ontological theory. 
The general theory of relativity suggests that one uses 
the "spacetime interpretation" of the spacetime concept 
presented below, although not even the general theory of 
relativity strictly speaking requires one to interpret 
the spacetime concept in that way. In the general theory 
of relativity, spacetime is considered to be something 
real, and it plays a central and active role in the 
physical course of events. Spacetime interacts with 
that which fills up spacetime, such as particles and 
fields. The curved geometry of spacetime dictates not 
only how a body is to move, but the body in turn 
dictates how spacetime should curve.\\
	
This other possible interpretation of the concept of 
spacetime, is that it can be understood as a merge 
of the two separate concepts space and time, 
into a single entity called \textit{spacetime}.

According to what has been said above, in the Newtonian 
world view, is what exists a three-dimensional (spatial) 
space at single point in time, \textit{the present}. Time is there 
absolute and is ticking at the same rate regardless of 
what is going on in the three-dimensional space. As time 
goes, one present is replaced by its subsequent present, 
which in turn is replaced with its subsequent present, and 
so on. Neither all those present moments which have been, 
the so-called past, nor all those present moments which 
not yet have come into existence, the so-called future, 
exist. Only the present moment exists. The present 
corresponds to a snapshot of a three-dimensional space, 
together with everything in this space at this moment 
in time. 

Some examples of things or events that can occur 
in the present, can, e.g., be a person on Mallorca who 
drops a coin, a bullet which leaves the muzzle of a rifle 
on the north pole, a meteroit colliding with a satellite 
above the earth surface, or a star exploding in another 
galaxy. All these events occur in an objective and 
absolute present. In the next moment on the absolute 
time, this present moment does not exist anymore, but 
instead belongs to the past. The past has no existence 
in itself, but is just a name on those present moments 
which previously have existed. In the same way, the 
future is not something that exists, but is just a 
name on those present moments which are yet to come. 
In the Newtonian approach, the past and the future are 
really only something which exist in relation to a 
brain existing in the present.
	
Spacetime is instead a four-dimensional space which 
exists. Three of these four dimensions constitute the 
usual three-dimensional (spatial) space and the 
remaining dimension is the usual time. Instead of 
reality being a three-dimensional present which is 
constantly being replaced by a new present as time goes, 
spacetime is instead a four-dimensional "present". 
Since time is included in the four-dimensional spacetime, 
there is no time \textit{in motion}. Instead the four-dimensional 
spacetime, together with all its contents (such as particles 
and fields), is something which only \textit{is}. Thus, it is not 
only what one in a Newtonian approach to reality calls 
the present which exists. Instead the past, present, and 
future all exist to the same degree. All three are a part 
of spacetime, and exist all at once. In fact, what we have 
called the past, present, and future in the Newtonian 
approach, have no direct counterparts, or absolute 
meaning, in spacetime.\\ 

This is what is meant by \textit{spacetime} in the "spacetime 
interpretation". But how can a consciousness and an 
experience of the passage of time arise from such a 
timeless and "frozen" spacetime? How can one reconcile 
the spacetime concept, interpreted according to the 
"spacetime interpretation", with the experience all 
of us have of a time in motion, a time that goes, and 
that the personal present is the only moment in time 
which exists for us, one personal present at a time?\\
 
If spacetime is real in the way described above in the 
"spacetime interpretation", then the only possibility 
I can see to reconcile, on the one hand, the timeless 
concept of spacetime with, on the other hand, the 
emergence of a consciousness, an experience of a time 
in motion and that we exist only in a single unique 
personal present, is that the latter is some sort of 
"illusion". ("Illusion" is not a really satisfactory 
word, but it is the best word I can come up with, that 
comes closest to describe and capture what I mean.) By 
the word "illusion", I do not mean that the spacetime 
events underlying the illusion are unreal. I mean that 
the personal experience that the brain creates of a 
time that \textit{goes}, a time in \textit{motion}, 
and a reality that only exists 
\textit{one moment at a time}, is an "illusion".\\

How, and why, our brains could or would be designed 
to operate in this way, is something I do not want 
to speculate too much about. But perhaps it could be 
a kind of "side effect" of a four-dimensionally 
existing brain? Perhaps some evolutionary advantage 
might lie behind? An evolutionary successful 
four-dimensional brain of our kind could perhaps 
more easily have arisen if this kind of "side 
effect" also arises? That some four-dimensional 
formations in spacetime have been equipped with 
the ability to create
an experience of a time in motion, that events have 
a time order, and a reality existing one moment at a 
time, etc, is perhaps something that gives these 
four-dimensional formations the right conditions and 
advantages needed for them to arise?\\

Bear in mind that we do not really have a completely 
clear picture, or understanding, of how thoughts, and 
an experience of a consciousness, can arise even if we 
were to adopt a Newtonian world view. We understand, or 
imagine, that thoughts, and consciousness, probably have 
to be a product or effect of the particles and fields 
that constitute our brains. But exactly how this 
experience or effect arises from these building blocks, 
is something that we only partly understand. So our 
inability in the present-day situation to understand 
how an experience of a consciousness and the passage 
of time, may arise from the timeless spacetime concept, 
is in itself no argument for rejecting a timeless 
interpretation and description of the concept of spacetime.\\ 

Let us simply assume that it is possible for an experience 
of a consciousness, and a "time in motion", to arise from 
such a timeless thing as spacetime. If one looks at how we 
perceive reality and our everyday world, it probably seems 
very far-fetched, speculative, and unlikely that it could 
be like this. But another consistent, 
systematic, and coherent way to understand the concept of 
spacetime on, I cannot see if one wants to get away from a 
notion of absolute simultaneity. So if this is 
the consequences of what is thought and worked out from 
a consistent, systematic, and logical reflection of 
reality as it appears to us humans, perhaps we need to 
find us to accept these consequences, although they may 
seem strange or unbelievable to us. That reality, seen 
from a larger perspective, may seem unintuitive, remarkable, 
and unlikely, is in itself not something strange, or an 
argument for that it would not be possible for it to be 
like that. Because reasonably our brains are of 
evolutionary reasons only constructed to understand the 
aspects of reality which constitutes our "closest or most 
immediate reality", or, put in another way, 
our everyday world.

\section{The various interpretations commented and compared}
\label{sec:varincomcomp}

Why is it really wrong to imagine reality in a Newtonian 
or Lorentzian way? The "spacetime interpretation" is 
perhaps more in line with what we strictly seen can say 
about reality. But does it not also move the description 
of reality in physics further away from our everyday
world, and thereby from an intuitive understanding and 
conception of reality? Cannot the assumption of an 
abstract absolute space be compensated by the fact that 
reality then becomes easier to grasp? It may be difficult 
to understand, and perhaps seem bold, that Einstein 
discarded an intuitive and comprehensible notion of space 
and time, and replaced it with an operational description 
of space and time. And this just to not be forced to
assume the existence of an absolute space or an ether, or? 

It is even more remarkable that Einstein did this 
without "going all the way" and coming up with the concept 
of spacetime. For it was only in and with the spacetime 
concept that I can see that one could give a satisfactory 
explanation, obtain an overall picture and a deeper 
understanding of the special theory of relativity and its 
predictions. Because the spacetime concept was not in the 
original version of the special theory of relativity as 
presented by Einstein. It came instead after that
Minkowski, who was Einstein's former teacher in 
mathematics, reworked Einstein's special theory of 
relativity to a different mathematical form (up to the 
point that not even Einstein for a while recognized his 
own theory). The spacetime concept was a product of this 
reworking. Initially Einstein disliked the concept of 
spacetime, but eventually took it to his heart. It was 
later to enable and become the foundation of his general 
theory of relativity.\\

So what is most satisfactory: 1) To understand how 
reality works at the price of the having to assume an 
abstract concept which has no direct support in 
observations, but which feels as intuitively obvious, 
and gives us an opportunity to visualize and understand 
reality based on everyday concepts and everyday thinking, 
i.e., by "common sense"? Or 2) to stop at an understanding 
and description of reality that only use concepts which
are defined on the basis of something that can be 
measured and observed?\\ 

These questions are related to what is really meant by 
\textit{to understand} something. Even if we understand something 
logically and operationally, this does not necessarily 
mean that we experience it as if we "truly understand" 
this something. Because, how do we know or decide that 
we really understand reality, if we cannot imagine and 
explain reality in terms of everyday concepts and 
everyday thinking? For what other criteria should we 
have on a theory, for us to truly think that it explains 
and describes reality in an ontological way? Or is perhaps 
the ambition to understand and describe how reality is 
objectively, a naive, unrealistic, and hopeless dream? 
Should one perhaps let go of this ambition and be content 
with trying to create theories that are as successful as 
possible when it comes to describe existing, and predict 
new, observations in an operational way? How one views and 
responds to these questions, have to do with one's attitude 
towards, and opinion of, what the role and final goal of 
science should be. The above questions are, e.g., related 
to how one looks at a theory such as quantum mechanics, 
which some claim gives us a \textit{complete} 
description of reality.\\ 
	
If alternative 2) above seemed obvious to the reader, then 
remember that, e.g., the atomic hypothesis has 
no, or at least originally had no, direct support in 
perception and observations. Originally the atomic 
hypothesis only indirectly had support in observations, 
since it predicted, described and explained, e.g., the 
behaviour of gases, the structure of the periodic table, 
Brownian motion, etc. To me approach 2) also has a 
sense over it of giving up too easily or too early.\\ 

But in practice, there will probably always be spokesmen and
representatives of both these two different alternative 
approaches and ways of working, among scientists and other 
thinkers. Not only in different individuals, but also in 
one and the same individual.\\

The "Lorentzian interpretation" offers an explanation of 
how the special theory of relativity fundamentally works. 
It explains how reality manages the feat of making the 
speed of light and the laws of nature to appear invariant, 
since time \textit{really} slows down and 
distances \textit{really} become 
shorter, for observers in motion relative to the absolute 
space. But this comes at the price of having to assume the 
existence of an absolute space which cannot be observed,
and this can be seen as a science-philosophical 
deficiency or weakness in this interpretation.\\

But perhaps there could be other ways to interpret the 
special theory of relativity, that do not involve 
an absolute space, ether, or a spacetime? It is, of course, 
not inconceivable that there could be. But as long as 
these ideas are the best we have, what else can we do 
other than to believe in and use these ideas?\\ 

However, 
now it is not in any way the case that the spacetime 
concept is a pure invention or something purely made up. 
It is a consequence of a logical, systematic, and
consistent scientific and philosophical thinking. A 
reasoning containing a minimum of assumptions that are 
not supported by observations. And how else can we build 
our theories about reality, if we do not want to get into 
metaphysics and speculations?\\ 

Moreover, if it really 
would be the case that there are no such things as an 
underlying absolute Newtonian space, or absolutely 
simultaneity, then we would be forced to accept that 
the Newtonian way to imagine space and time would be 
incorrect. And if we have given up this idea, what 
other choice do we then have, other than to create a 
conception of space and time from what we \textit{really} 
know about reality, i.e., what we can observe?\\ 

Now the criticism of the spacetime concept would 
perhaps be more justified, if it had not turned out 
to be successful in some other way than in the special 
theory of relativity. But the spacetime concept 
enabled and led to the general theory of relativity, 
which is a successful, deep and conceptually satisfactory theory. 
It contains Newton's law of gravity as a special case, 
and in the cases it has been tested it has also been 
confirmed by observations. The theory has led to an 
increased understanding, and given rise to new theories, 
of reality. And is not being successful, strictly seen, 
the only criterion one can have on the correctness of a 
theory? At least from an evolutionary perspective this 
seems to be the case. 

Then there are, of course, many 
other guiding principles, and criteria, on a theory that 
enter. E.g., that it should contain a minimum of unfounded 
assumptions and concepts that do not find support in 
observations. It also should, as far as possible, be 
consistent with other already existing theories. But 
preferably also with our everyday thinking, for it is 
when a theory is this, that we feel that we really 
understand it and we can visualize reality.\\ 
	
In the future it may, of course, turn out that the 
spacetime concept ceases to be successful and is 
therefore not quite right. It is not at all 
inconceivable that the idea of an ether, or an absolute 
space, could make a come back. However, the tendency 
is rather the opposite, if one considers how the 
theories of physics look like in the present-day 
situation. The concept of spacetime is 
certainly accepted and taken to be correct by the 
physics community. At the same time, it seems to me that the 
spacetime concept not always is being interpreted in 
the manner I have described above in the "spacetime 
interpretation". Many still seem to view space and 
time in a more absolute Newtonian way, even though 
they know about the theory of relativity and the 
concept of spacetime.\\ 

My impression is that the more 
fundamental research area one looks at, the more is 
spacetime a more known, recognized, embraced, and 
used concept among physicists in this field. I am 
here thinking of such areas as, e.g., particle 
physics, astrophysics, and string theory. However, 
when it comes to the scientific community at large, 
my impression is, at the same time, that the concept 
of spacetime is unclear to many. Or that they do not 
consider spacetime as something real, but rather a 
useful abstract concept. The majority are aware of 
the consequences of the special theory of relativity, 
such as time dilation and length contraction, and 
they know about the concept of spacetime. But not 
many seem to have a deeper insight, or understanding 
of the special theory of relativity, or the concept 
of spacetime. Or is it perhaps the lack of same from 
my side, that makes me experience the situation 
in this way?

\section{Spacetime diagrams}
\label{sec:sptdia}

A spacetime diagram is a good way to visualize and 
understand the special theory of relativity and the 
spacetime concept on. It is enough to consider the 
case of a two-dimensional spacetime, i.e., with one 
space dimension and one time dimension. It is easy to 
generalize this to the case when one has two or three 
space dimensions. Figure~8 shows an example of 
a spacetime diagram.\\

\begin{figure}
\includegraphics[scale=.8]{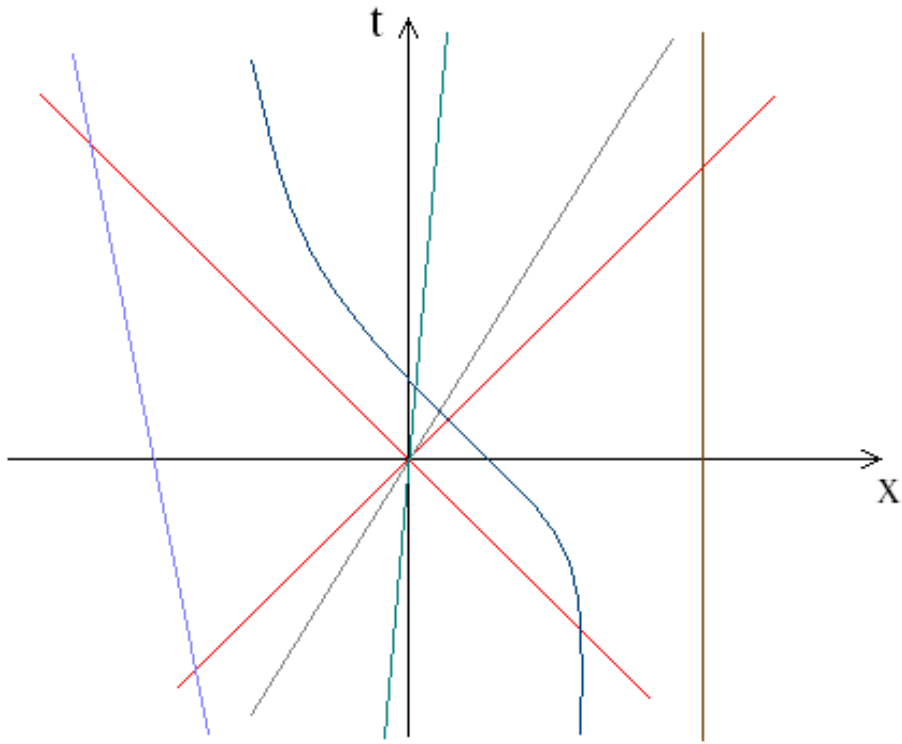}\\
\caption{An example of a spacetime diagram.}
\end{figure}

The space dimension $x$ is on the horizontal axis and 
the time dimension $t$ on the vertical axis. Each of the 
lines in the spacetime diagram represents the motion of 
an object, such as a particle, in space with respect to 
time. The lines describe where each object in space are at 
every point in time. The slope of the curve gives the 
object's speed. If a curve is not straight, this means 
that the speed is not constant, but the object instead 
describes an accelerated motion. In this case, the 
instantaneous velocity of the object is given by the 
slope of the tangent to the curve.\\ 

The red lines show 
the motion of two light beams in space, which occur at 
the highest possible speed, i.e, the speed of light, 
approximately 300 000 000 m/s. For simplicity we have in 
figure~8 chosen to use a second as time unit, 
and as length unit a \textit{light second}
(ls), which is 300 000 000 m. Light speed expressed in 
light seconds is then 1 ls/s. All other speeds have a 
value between 0 and 1 ls/s, where the speed 0 means that 
the body is at rest. Light beams then always move according to the 
special theory of relativity along straight lines 
making an angle of 45 degrees to the coordinate axes.\\

Note that spacetime is not an Euclidean space where 
distances $s$ are given by the Pythagorean theorem, 
i.e., $s^2 =  t^2 + x^2$. In spacetime distances are 
instead mathematically defined by "$s^2$" $= t^2 - x^2$. 
(We could just as well have chosen to define
distance as "$s^2$" $= x^2 - t^2$, i.e., with opposite sign.) 
Since this distance also can become negative, it cannot 
in general be seen as the square of a distance $s$; hence 
the quotes around $s^2$ above. One can instead consider 
"$s^2$" as a single symbol. But this in itself is nothing 
strange or mysterious. We can choose ourselves to define 
distance in spacetime as it pleases us, as long as we do 
it in a consistent manner so that no logical 
contradictions arise. That we have chosen to denote the 
distance measure as the square of something is just to 
show an analogy with the Pythagorean theorem. Since I do 
not want to go too much into the mathematics here, it is 
enough to explain in this rough or sketchy way.\\

Each point in the spacetime diagram represent a unique 
\textit{event} in spacetime. The two-dimensional plane therefore 
consists of all events that take place in spacetime. The 
coordinates $x$ and $t$ correspond to the position and time 
coordinates, respectively, relative to a reference body 
or inertial reference system which we have chosen to 
regard as being at rest. Each point in space on this 
reference body has a position coordinate $x$ equal to the 
number of measuring rods away the point is from our 
chosen origin in space. At every point in space we can, in 
principle, imagine that there is a clock which is 
synchronized with all other clocks that are at rest 
relative to this reference body. Every point in spacetime 
can therefore be indicated by a unique coordinate pair 
$(x, t)$, which gives the spacetime coordinates for this 
event relative to the reference body that we have chosen 
to regard as being at rest.\\ 

Since there are no reference 
bodies which are in absolute rest, we could just as well 
indicate each point in spacetime relative to another 
reference body with coordinates $(x', t')$. There are 
infinitely many inertial reference systems; more 
specifically, one for each velocity between -1 and 1, 
indicating the relative velocity of a reference body 
relative to a specific, but arbitrarily selected, 
reference body, e.g., the reference body with coordinates 
$x$ and $t$. (Inertial reference systems which differ only by 
a normal rotation in the spatial space, I here regard as 
equivalent, and not as different inertial reference 
systems.) A negative velocity indicates that a body is 
moving in the negative $x$-direction in space. Figure~9 
shows where the coordinate axes, belonging to the 
reference body with coordinates $x'$ and $t'$, will be in 
relation to the reference body with coordinates $x$ and $t$.\\

\begin{figure}
\includegraphics[scale=.8]{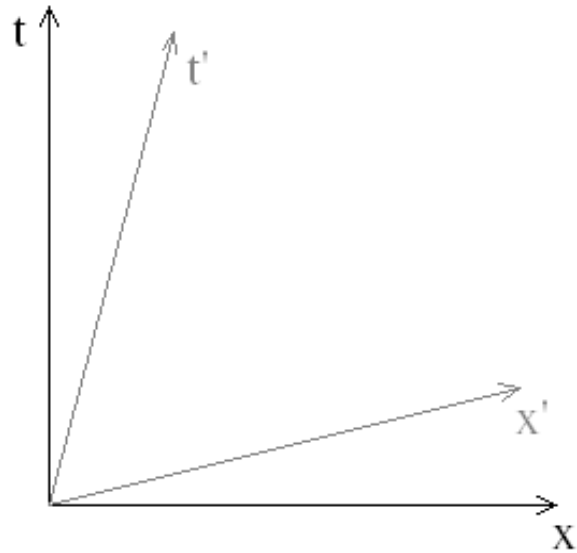}\\
\caption{The coordinate axes of two different 
inertial reference systems shown in relation to one another.}
\end{figure}
			 
Let us refer to the coordinate system $(x, t$) as $K$, and 
the corresponding reference body as $R$; and similarly for 
the other coordinate systems and reference bodies. We have 
here chosen to consider $R$ 
as being at rest. Therefore the $t$-coordinate axis points 
only in the time direction of $K$. Since $R'$ moves relative 
to $R$, the clock in the origin of $K'$ moves relative to $R$ 
along the line that constitutes the $t'$-axis in the 
coordinate system $K$.\\ 

A line which is parallel with the space axis of a
coordinate system, corresponds to events that are 
simultaneous relative to one another in this coordinate 
system. E.g., are all the clocks that are at rest 
relative to $R'$ and are located along the $x'$-axis, 
synchronized to one another relative to the coordinate 
system $K'$, and all show the time $t' = 0$ (see 
figure~9).\\

Consider different reference bodies which are in 
constant uniform motion relative to one another, 
i.e., different inertial reference systems. A 
coordinate system is thus a set of coordinate 
pairs $(x, t)$, which in a systematic way indicate 
all the events in spacetime relative to a certain 
inertial reference system. For each coordinate system, 
there is to each event a corresponding coordinate pair, 
i.e., each event corresponds to an infinite number of 
different coordinate pairs.\\

Inversely, assume that one has labeled each event in 
spacetime with a coordinate pair $(x, t)$. All these 
coordinate pairs together constitute a coordinate 
system. Further assume that one has labeled the events 
in spacetime in such a way, that one has an infinite 
number of different such coordinate systems. This means 
that each event is labeled by an infinite number of 
coordinate pairs. In addition to this, also assume 
that one has labeled the events in such a way, that 
these coordinate systems are related to one another 
in the way that the special theory of relativity 
(Lorentz equations) prescribes that they should. 

This gives us a set of an infinite number of different 
coordinate systems, which in a systematic way label 
and structure all the events in spacetime. One can 
now choose to view each coordinate system as 
corresponding to an inertial reference system. The 
space and time coordinates in every coordinate system, 
then correspond to the position and time of events 
relative to the corresponding inertial reference system. 
Hence can, for instance, the set of events in spacetime 
corresponding to straight lines parallel with the 
space axis of any coordinate system, be viewed as simultaneous 
events in the inertial reference system corresponding to this 
coordinate system. And the set of events 
corresponding to straight lines parallel with the time 
axis of any coordinate system, will be the paths in 
spacetime that objects, which are at rest relative to 
some reference body, will follow (see figure~10).\\

\begin{figure}
\includegraphics[scale=.8]{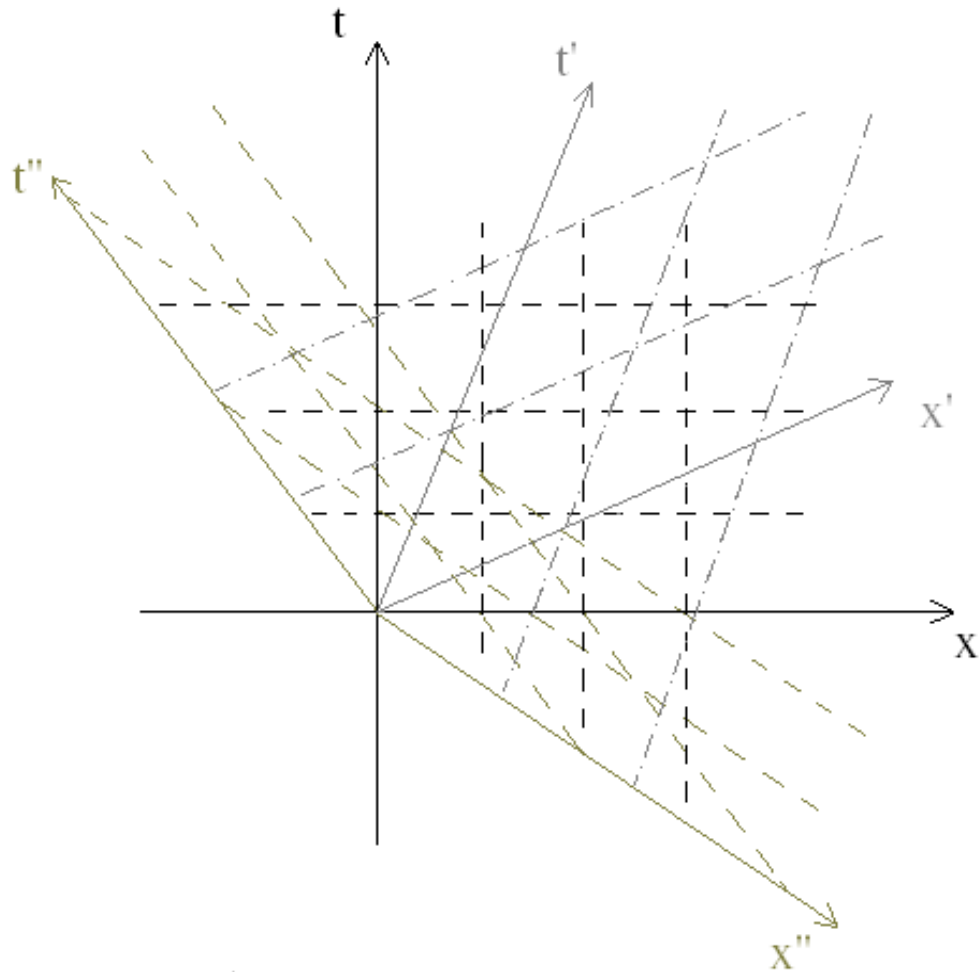}\\
\caption{Coordinate systems of different inertial
  reference systems shown relative to one another.}
\end{figure}
		 
One can think of a spacetime diagram as a graphical 
representation of the Lorentz equations. By studying a 
spacetime diagram, one sees how space, time, and 
simultaneity change for reference bodies in motion 
relative to a reference body at rest. Thanks to the 
spacetime diagram representation of spacetime, one can 
demonstrate both the Lorentzian and the spacetime 
interpretation of the special theory of relativity 
only by interpreting the diagram in two different ways.\\
 
The Lorentzian interpretation, which is of absolute 
character, means that \textit{one and only one} of the coordinate 
systems in the spacetime diagram corresponds to a
reference body which is in absolute rest relative to the 
absolute Newtonian space. Assume that the reference body corresponding 
to the $(x, t)$ coordinate system, is in absolute rest. In the Lorentzian
interpretation, time is also absolute. Assume that in the 
present moment in time, the absolute time has a certain 
value, say $t = 0$. This means that only those events that 
are on the $x$-axis in figure~10 exist. All other 
events in the spacetime diagram do not exist in this 
moment in time. All other reference bodies are only 
\textit{seemingly} (i.e., 
seen from an observer's point of view) equivalent to 
this absolute reference body. Thus, for all other 
reference bodies, time \textit{objectively seen} goes slower and 
distances \textit{objectively seen} are shorter, relative to 
the reference body which is in absolute rest. By studying 
the spacetime diagram with this in mind, it is easier to 
understand how, according to the Lorentzian interpretation, 
all reference bodies can seem to be equivalent.\\
 
With a spacetime interpretation, there is no inertial 
reference system which is in absolute rest. All events 
in spacetime have the same degree of existence. One 
could compare the spacetime with a tabletop (which we can 
imagine to be so large that we do not see the edges of the 
tabletop). Imagine that one has drawn up straight lines 
across the whole tabletop, all parallel with one another. 
Now it is not the case that the tabletop only exists along 
\textit{one} of these lines, or that the tabletop only exists along 
one of these lines \textit{at a time}. For what would it even mean 
to say such a thing!? The tabletop is, of course,
something we imagine exists as a whole. 

It is the same way with 
space and time in the spacetime interpretation. Space is 
there not something which exists along one parallel line 
at a time in spacetime, as the hand of the absolute time 
moves around its clock face. ["One parallel line at a
time" would instead have become "one parallel plane at a 
time", if we would have considered two space dimensions 
instead of one space dimension, and "one (parallel) 
three-dimensional space at a time" if we instead would 
have considered three space dimensions.] Spacetime is 
not a (infinite) three-dimensional space that, together
with its contents (such as particles, magnetic 
fields, stars, planets, etc), change appearance 
"as the hand of the absolute time \textit{moves} 
around its clock face". For the first, there is no
absolute time according to the spacetime interpretation. 
Secondly, spacetime is not something that \textit{changes with 
time}, since time is already included in spacetime. 
Spacetime only \textit{is}. Spacetime exists as a whole. 
Different inertial reference systems or coordinate systems, 
are just different ways to arrange ("to network", draw up, 
or structure) spacetime. All inertial reference systems are 
equivalent. No inertial reference system can claim to be 
"absolute" or "more correct" than any other inertial 
reference system; in the same way as no coordinate system 
(with orthogonal axes) on a plane, can be considered to be 
"absolute" or "more correct" than any other coordinate 
system (with orthogonal axes) [see figure~11].\\

\begin{figure}
\includegraphics[scale=.65]{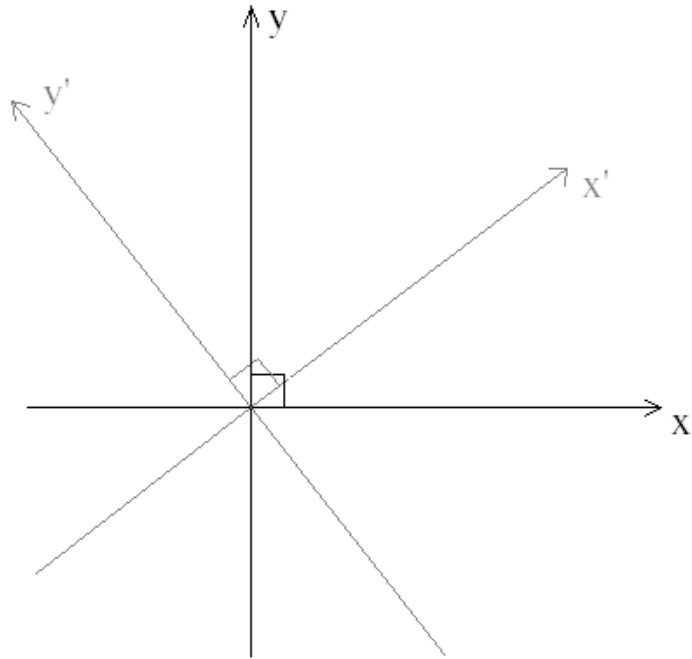}\\
\caption{Two different space coordinate systems of the
  same plane.}
\end{figure}
		 
According to what has been said above, a coordinate system 
$K$ in spacetime allows the concept of inertial reference 
system, reference body $R$, and the concept of an observer 
who is at rest relative to $R$. Each and every one of all 
the lines that are parallel to the $x$-axis of $K$, corresponds 
to $R$ at different points in time $t$. All events that are 
on such a line, occur simultaneously relative to $R$ at a 
time $t$ (given by the intersection of this line and $t$-axis of $K$).\\
 
The time order between two events $A$ and $B$ in spacetime, i.e., 
which one that occurs before or after the other one, is 
coordinate system dependent. If $A$ and $B$ are simultaneous 
relative to a reference body $R$, then it is always possible 
to find a reference body $R'$ where $A$ occurs before $B$, and a 
reference body $R''$ where $B$ occurs before $A$. Hence the time 
order is not absolute. To ask which of the events $A$ and $B$ 
that \textit{really} occurs first, makes as little sense as to ask 
which of two points in a plane that lies "highest up" in
the plane. Which point that lies "highest up" depends on 
which coordinate system one refers to.

\begin{figure}
\includegraphics[scale=.8]{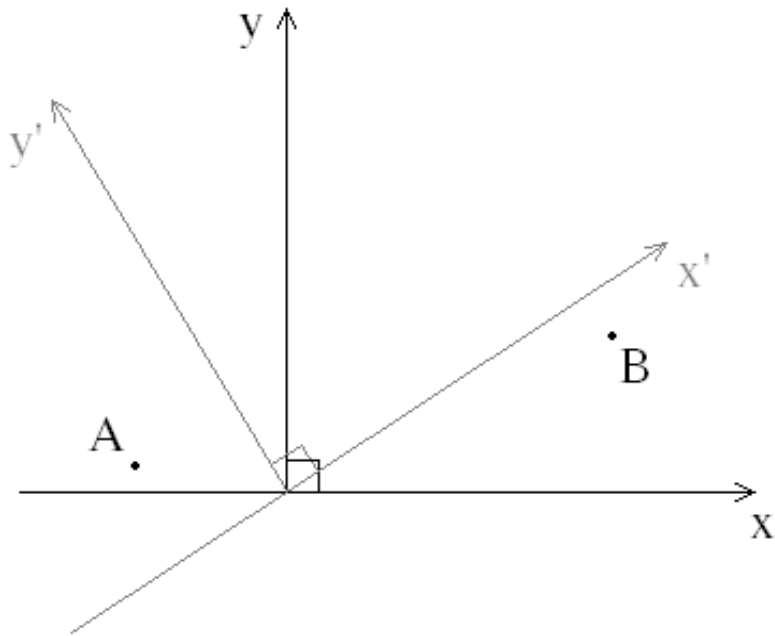}\\
\caption{Two different space coordinate systems.}
\end{figure}

\begin{figure}
\begin{center}
\includegraphics[scale=.8]{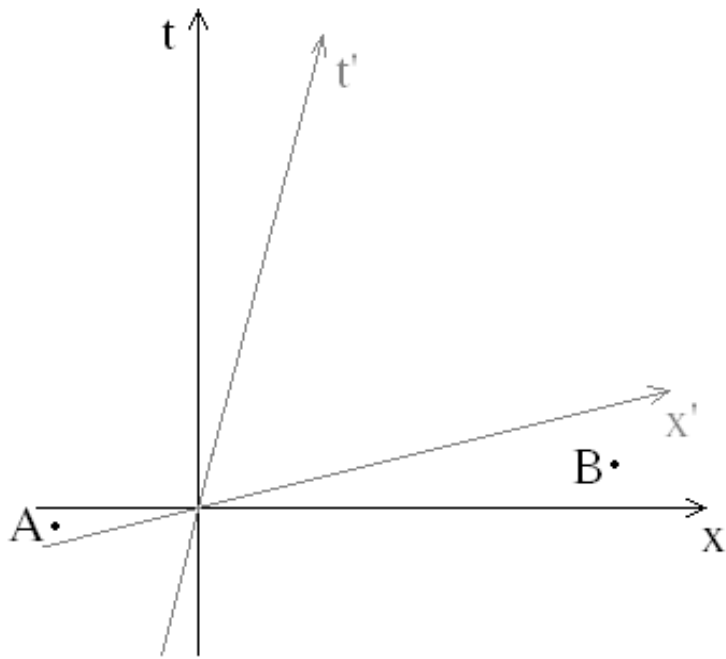}\\
\caption{Two different spacetime coordinate systems.}
\end{center}
\end{figure}
	
Consider two different coordinate systems $K$ and $K'$, with 
orthogonal coordinate axes, in a two-dimensional spatial 
space (see figure~12). Which of the 
points $A$ and $B$ that lies "highest up" in the plane, is 
coordinate system dependent. In $K$ point $B$ lies higher up 
than $A$, because it has a larger $y$-value. But in $K'$ 
point $A$ lies instead higher up than $B$, because it has a 
larger $y'$-value. It is therefore not meaningful to ask 
which of two points that lies "highest up" in a plane 
in any absolute sense, but it only makes sense if one 
also specify the coordinate system that one refers 
them to. Analogous to this, the time order between 
events $A$ and $B$ (see figure~13) depends on which 
coordinate system one compares them in. 
Relative to $K$ the event $A$ occurs before $B$, 
while relative to $K'$ the event $B$ occurs before 
$A$ (since $B$ lies below and $A$ above the $x'$-axis).\\
	
However, in spacetime there exist (infinitely many) 
combinations of pair of events $A$ and $B$, with the same 
time order relative to one another in \textit{all} 
inertial reference systems. This is the case for all events 
which are in each other's \textit{future} or 
\textit{past light cones}. An event's 
\textit{light cone} is defined as, all the events which 
can reach, or be reached from, this event with a signal 
whose maximum speed is the speed of light. To every event 
in spacetime there is a unique light cone (see figure~14). All 
events that are within the light gray area of spacetime, 
belong to event $O$:s \textit{future light cone}, while all 
events in the dark gray area belong to $O$:s \textit{past light 
cone}. All events that belong neither to the past, or 
future light cone to an event, are said to lie \textit{elsewhere}. 
Each event which is located elsewhere relative to $O$, is 
in some inertial reference system simultaneous with $O$. 
These different areas are bounded by lines (or cones in 
a three-dimensional spacetime) corresponding to the 
paths of light beams through spacetime.\\

On the other hand, for such pairs of events $A$ and $B$, 
one can also always 
find an event $C$ which both $A$ and $B$ are simultaneous 
with in the following sense: Suppose that the event $B$ 
lies in the event $A$:s future light cone. Then there is 
always some event $C$, which in some coordinate system $K'$ 
is simultaneous with the event $A$, and in another coordinate 
system $K''$ is simultaneous with $B$ (see figure~14). 
An event $C$ can thus be simultaneous with two events $A$ and 
$B$ having the same time order in all inertial reference 
systems.\\

It is therefore perhaps tempting to regard
events as more or less simultaneous, depending on whether 
they lie in, or outside one another's light cones. So 
perhaps it is, in some sense, meaningful to speak 
about an absolute, objective or ontological time order 
between certain events in spacetime after all?\\

\begin{figure}
\includegraphics[scale=.6]{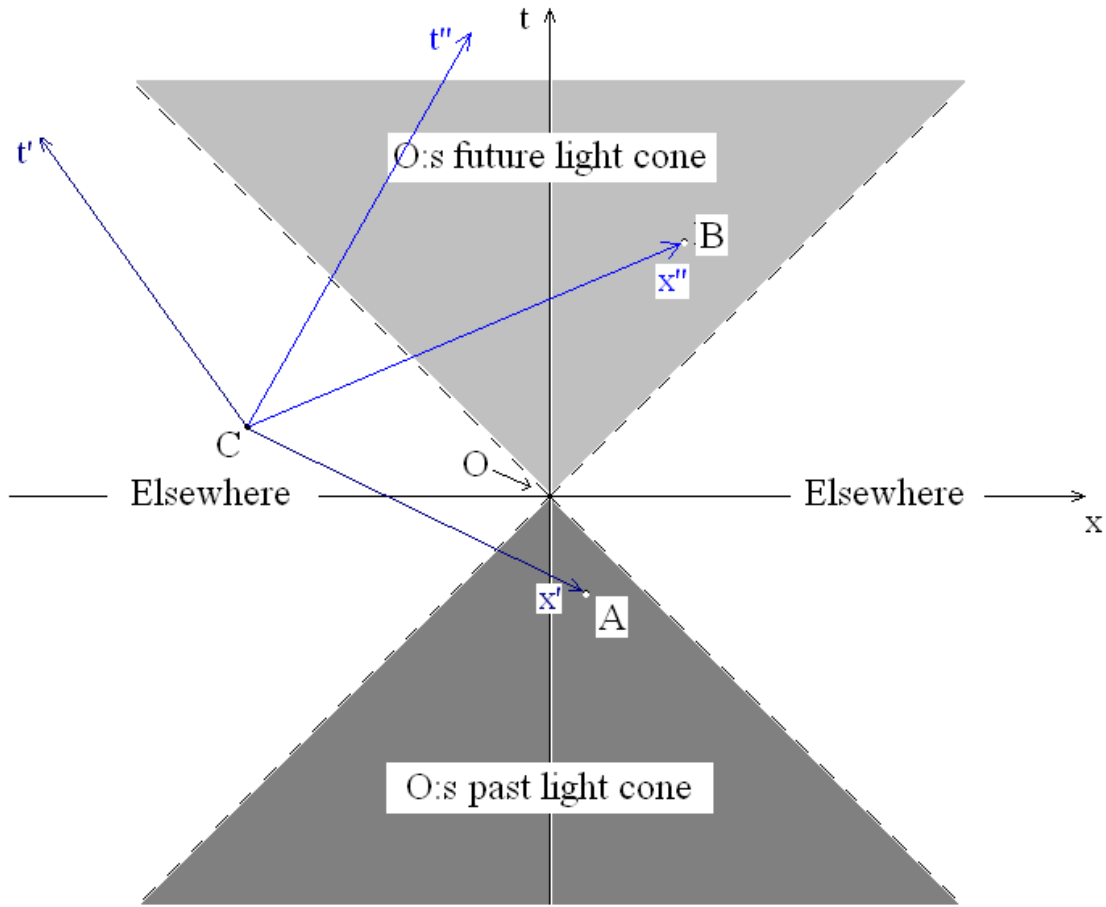}\\
\caption{The light cone to an event O in spacetime.}
\end{figure}

However, the fact of the matter is that, in and with the 
special theory of relativity, Einstein replaced the notion 
of absolute simultaneity with a \textit{definition} 
of the concept of simultaneity. 
Although a natural definition, it is strictly seen 
still an arbitrary definition of what is to be meant by 
the concept of simultaneity. A definition used 
by observers in the operational special 
theory of relativity as well as in the Lorentzian and 
spacetime interpretation. However, the Lorentzian interpretation 
assumes in addition to this that absolute simultaneity 
objectively seen exists. In the spacetime interpretation, 
spacetime is instead something which objectively seen 
exists as a whole. 
Those parts of spacetime which are "more towards the upper 
future part" of the spacetime plane, therefore cannot be 
considered to be more real than those parts of spacetime 
which are "more towards the lower past part" of the 
spacetime plane.\\

So concepts such as \textit{simultaneity}, 
\textit{past} and \textit{future light cone}, 
\textit{elsewhere}, etc, are created by and exist in 
relation to observers. What objective or ontological counterparts 
and significance these concepts have, depend on the 
interpretation one makes of the special theory of 
relativity. Because operationally or strictly seen
these concepts only show different 
conditions, relationships or correlations 
that exist between events. E.g., that an 
event $B$ lies in another event $A$:s future light cone, 
shows that event $B$ can be reached from event $A$ with a 
signal travelling at a speed that is lower than the 
speed of light. Or if the events $A$ and $B$ instead would 
be simultaneous in some inertial reference system, 
this shows, e.g., that it is not possible to send a 
signal between the events with a speed that is lower 
than the speed of light.\\ 

But what about the experience we all have that time 
has a direction? Because we experience that time goes 
from the past towards the future, and that we can 
influence what will happen in the future, but not what 
have happened in the past. It is true that time 
and simultaneity are inertial reference system 
dependent concepts in the special theory of relativity, 
but the direction of time is the same for all 
inertial reference systems, since time in all inertial 
reference systems points from a "common past" towards 
a "common future". Or to put it differently, the time 
axis in all inertial reference systems points "upwards" 
in a spacetime diagram. So in what way would then a time 
direction in spacetime really have an absolute meaning, 
if now spacetime and all its events exist as a whole?\\

With a spacetime interpretation, the time direction only 
has an objective or ontological counterpart and 
significance in that it shows 
certain conditions, relationships or correlations that 
exist between events in spacetime. The subjective 
experience of a time in motion and that time has a 
direction, is something that only exists in relation to 
a human brain or a (biological) machine of some other kind.

\section{The twin paradox}
\label{sec:twinpx}
 
We will now see how one can explain the so-called "twin 
paradox" according to a spacetime interpretation of the 
special theory of relativity. The reason that I choose the 
twin paradox as an example, is that I think it involves
much of that which is central, interesting and puzzling 
with the special theory of relativity.\\

The scenario of the twin paradox is the following: Imagine 
two identical twins that are on Earth. One of the twins 
travels into space on a spaceship at a speed close to the 
speed of light, while the other twin remains on Earth. After 
a number of years have passed on Earth, the twin travelling 
in space returns from his (or her) trip. But when the
twins are reunited, they are not of the same age anymore. 
The twin who stayed on Earth turns out to be older than
the twin who traveled in space. It could, e.g., have been 
the case that the space travelling twin aged 5 years, while 
the twin who stayed on Earth aged 9 years, i.e., the space 
travelling twin after his journey is 4 years younger than 
his twin brother(!).\\

But if motion is something relative, should not the 
situation be symmetrical for the two twins? Relative to 
the space ship, was it not Earth which instead moved and 
the spaceship which was at rest? Should not the time then 
have gone slower on Earth than on the space ship? And if 
so, should not the twin on Earth have been the younger of 
the two when they reunited? This seemingly paradoxical 
scenario is called the "twin paradox".\\

Let us first point out that it is an experimental fact 
that reality behaves as described above in the twin 
paradox scenario. Admittedly one has not managed to build 
any spaceship that can travel at a speed close to the 
speed of light, so the twin paradox has not been 
experimentally verified exactly in the way that the 
scenario above describes. But one has shown that the twin 
paradox is in agreement with what experiments show when it 
comes to the microscopic world, i.e., for microscopic 
particles such as elementary particles and atoms. And
there is no reason to expect that it would be different on 
a macroscopic scale, e.g., for space travelling twins.\\

To be somewhat poetic, if one wants to see more of the 
world, i.e., travel far and wide in space, one gets to 
see a lot in a short amount of time. But if one wants 
more time for reflection, one would be wiser to stay at 
home. However, this is a truth with modifications, 
because one's personal or subjective time always goes at 
the same rate regardless of how one moves. I.e.,
personally one does not experience that life becomes 
longer or shorter, no matter how fast one moves. However, 
different people's lives may seem shorter or longer in 
relation to each other, depending on how they move 
relative to one another.\\ 

The so-called "twin paradox" is however not a real
paradox. The special theory of relativity gives a 
completely logical and consistent explanation of 
why it must be the twin who went off on the spaceship 
that will be the younger of the two. The explanation 
in short has to do with the fact that the situation was 
not symmetric for the two twins. The difference between 
their situations is that the twin on Earth the whole time 
remained at rest relative to \textit{one and the same} inertial 
reference system, while the twin who was on board the 
space ship \textit{did not}. The twin on board the space ship in 
fact changed inertial reference system (at least once) 
during his trip. Because the spaceship must at some stage 
have turned around, in order to be able to return to
Earth. And this means that the twin on board the space 
ship was not at rest relative to one and the same 
inertial reference system during the whole trip.\\ 

It does not really matter for the "twin paradox" 
exactly how the space ship moved during the trip. 
The only important thing is that the twin at some stage 
returns to Earth, so that the twins' (biological) clocks, 
or in other words their ages, can be compared with one 
another when both are in the same place in space again.\\

Note that we in the train example above have already 
seen that the situation is completely symmetrical for 
both inertial reference systems when clocks instead are 
compared with one another at different places in space. 
But in the train example, the observers never changed 
their state of motion. Separately both of them remained 
the whole time at rest relative to one and the same 
inertial reference system. The embankment and the train 
never returned to a position that they at an earlier 
stage had already been in relative to one another. So in 
that case one was forced to make use of \textit{two} different 
clocks along the embankment to compare times on the two 
reference bodies. This meant that one had to make use of 
synchronized clocks. The relativity of simultaneity thus 
enabled both reference bodies to maintain their opinions 
that the clocks of the other reference body go more slowly.\\ 
	
Assume that the twin who went off on the space ship, did
it first with a constant speed away from Earth, and then 
returned with another constant speed (of course in the 
opposite direction since the space ship intends to return 
to Earth). How much time the space ship takes to
accelerate up to the constant speed it then maintains, 
does not matter for the demonstration of the twin
paradox. Nor does it matter how much time the space ship 
takes to reverse its motion, which also must involve an 
acceleration. One could even imagine that the change in 
velocity was instantaneous, i.e., a discontinous velocity 
change, or an infinite acceleration if you like. In fact, 
the twin paradox has nothing to do with acceleration. Nor 
has the twin paradox in principle anything to do with
twins for that matter. Instead to let a twin go off into 
space, one could instead replace the two twins with
clocks. Neither do we need to accelerate the spaceship to 
a certain velocity. Instead we assume that the space ship 
going off, instead just passes Earth with uniform constant 
speed. On board the space ship is a clock $B$. When the
space ship passes a clock $A$ on Earth, both clocks are 
set to show the time 0. The space ship then continues its 
journey into space with a constant speed.\\ 

At some point in 
time on the clock $B$ on board the spaceship, we assume that 
spaceship meets another spaceship which is moving towards 
Earth, also with a constant speed. On board this second 
spaceship is a clock $C$. We assume that the clocks $A$, $B$ and 
$C$ are all of identical construction, i.e., at rest they
are all "ticking at the same rate". When the space ships 
pass one another, the clock $C$ is set to show the time that 
clock $B$ shows in this moment. Clock $C$ thereby "takes over 
the time" from clock $B$, without clock $B$ itself needs to 
change its state of motion (inertial reference system). 
Clock $C$ then continues with constant speed back to Earth. 
When it passes clock $A$, one compares the times on the two 
clocks. One will then discover that clock $C$ shows a time 
which is before the time on clock $A$. In other words, 
clocks $B$ and $C$ in total appear to have been ticking at 
slower rate than clock $A$.\\

Note that we have not involved acceleration in the
picture, which means that the twin paradox is independent 
of acceleration. \textit{That} clock $C$ 
has "lagged behind" clock $A$ 
has nothing to do with how the space ships are moving. 
However, \textit{how much} clock $C$ has "lagged behind" clock $A$, 
depends on how the space ships have been moving; because 
the faster something is moving, the slower time goes.\\

Thus, the difference between the Earth clock's path and 
the spaceship clocks' path through spacetime, is that 
the path of the Earth clock involves \textit{only one} inertial 
reference system, while the path of the spaceship clocks 
involves \textit{two different} inertial reference systems. 
Figure~15 shows how the twin paradox can be illustrated 
in a spacetime diagram. One can clearly see how the 
"spaceship path" through spacetime involves \textit{two different} 
inertial reference systems, while the "Earth based path" 
through spacetime involves \textit{only one} inertial reference
system.\\

\begin{figure}
\includegraphics[scale=.8]{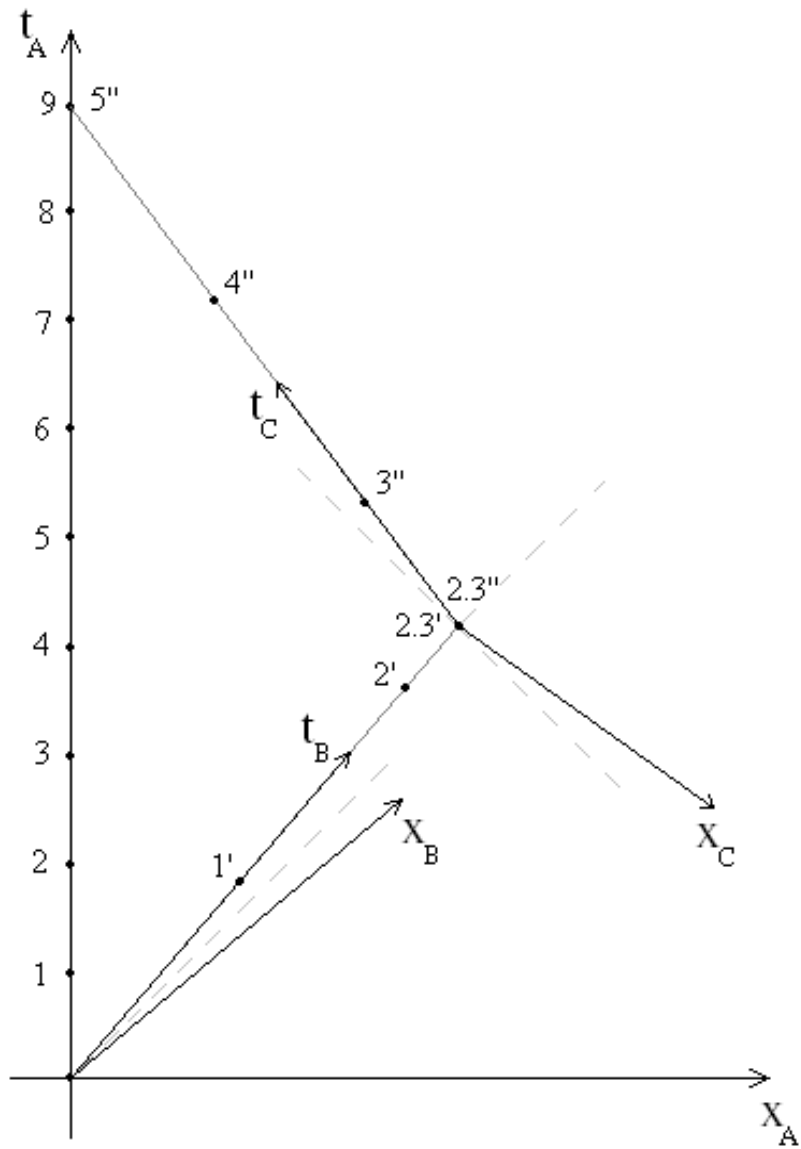}\\
\caption{A twin paradox scenario shown in a spacetime diagram.}
\end{figure}
		 
In the spacetime diagram shown in figure~15 one sees that the 
different paths form a triangle in spacetime. (Depending on how 
the involved spaceships move, the paths through spacetime 
could, of course, form a more complex geometrical figure 
than a triangle.) An interesting observation one can make, 
is that there is an analogy here with the \textit{triangle
inequality} in the ordinary spatial space. The triangle 
inequality basically means that the \textit{shortest} path through 
space, from a point $Q$ to another point $S$, is along a 
straight line. This means that, if one instead goes via 
a third point $R$, then one goes a \textit{longer way} (see 
figure~16). Analogous with this (though the other 
way around), in spacetime a straight line represents 
the \textit{longest way in time} between two points, whereas 
all other paths are \textit{shorter in time}.\\

\begin{figure}
\includegraphics[scale=.8]{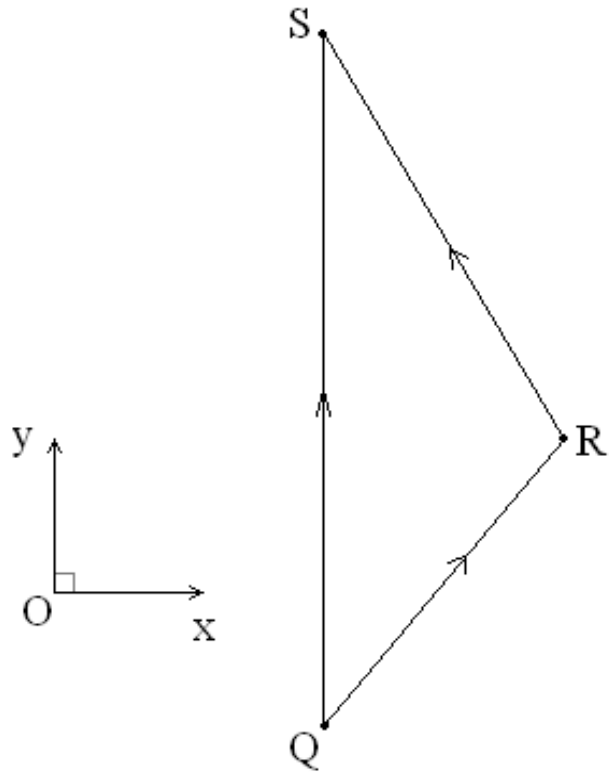}\\
\caption{The shortest path between two points in space is
  along a straight line.}
\end{figure}
		 		 
All inertial reference systems will agree on that it 
was the twin who traveled in space, and then came 
back again, who aged the least. All in accordance with 
the predictions of the special theory of relativity. 
The special theory of relativity does not contain any 
logical contradictions in the case of the twin paradox; 
or in any other cases for that matter. Consequently 
there is no real paradox, i.e., no twin paradox.\\

But how does then reality manage the feat to get \textit{all} 
inertial reference systems to observe that time moves 
slower for \textit{all} other inertial reference systems in 
motion relative to them? And, despite this, how does 
reality manage the feat to get \textit{all} inertial reference 
systems to agree on for whom time really has gone the 
slowest, when they decide to meet to examine the 
matter more closely?\\

The answers to these questions depends on which 
interpretation one makes of the special theory of 
relativity. Strictly seen, the special theory of 
relativity is an operational theory and it does not 
\textit{explain} why or how reality manages this feat. It only 
says that it logically must be in this way, if reality 
is such that the speed of light is the same and that 
reality appears the same for all inertial reference 
systems. The predictions of the special theory of 
relativity are logical consequences of these two 
postulates. And both postulates and the predictions 
of the special theory of relativity, have been 
confirmed by countless number of observations and 
experiments. The special theory of relativity is a 
logically consistent theory, that describes how space 
and time change for bodies in motion.\\ 

However, the "Lorentzian interpretation" and the 
"spacetime interpretation" of the special theory of 
relativity, both offer an explanation of how reality 
manages this feat. According to the "Lorentzian 
interpretation", reality manages this feat by making 
time \textit{objectively seen} go slower for clocks 
which move relative to the absolute space.\\

In the "spacetime interpretation", it is not in the same 
way meaningful to talk about who is objectively seen the 
youngest or oldest, or for whom time objectively seen goes 
slower. In general one needs to meet at an event in 
spacetime to settle the matter. To be able to answer 
questions such as "who is the oldest or youngest" and 
"for whom the time goes the slowest", when the
observers are at different events in spacetime, then one 
also has to specify relative to which inertial reference 
system one compares their ages or times to. And as we 
previously have established, are time and simultaneity 
coordinate dependent concepts. Therefore, it is not 
meaningful to ask which of two events that really or 
objectively seen occur before or after another event. All 
events in spacetime exist in the same way, and "all at 
once". Spacetime \textit{is}.

The coordinate systems of the different inertial reference 
systems, just correspond to different ways to organize 
events in spacetime on, and they give the \textit{geometry} and 
\textit{metric} of spacetime. The grid of a coordinate system in 
spacetime thus shows how events are related to one another 
and gives the distances between events in spacetime. As we 
previously have pointed out, by distance is not meant an 
Euclidean distance, but distance in a spacetime sense, 
i.e., "$s^2$" $= t^2 - x^2$. One can show that it follows 
from the Lorentz equations, that this distance measure 
is independent of the coordinate system it is measured 
in and thus works as an invariant distance measure in 
spacetime. All observers agree on what this spacetime 
distance measure between two events is.\\

From the spacetime diagram in figure~15, one sees that the 
spacetime distance between the event that the spaceship 
left Earth and the event that it passed clock $C$, is 2.3 
years ("$s^2$" $= 2.3^2 - 0^2 \Rightarrow$ "$s$" 
$= 2.3$ years). This 
spacetime distance involves only a change in time, because 
relative to clock $B$:s inertial reference system, both 
events occur at the same place in space. There is only 
one inertial reference system for which this is the case. 
For all other inertial reference systems the events occur 
at different places in space, and then the space distance 
$x$ also enters the spacetime distance measure. The
spacetime distance between the events, thus becomes the 
distance in time on clock $B$ between the events, also 
called the \textit{rest time} or \textit{proper time}, 
which in this case is 2.3 years. 
In the same way, the spacetime distance between the event 
that clock $B$ passes clock $C$ and the event that clock $C$ 
reaches Earth, is 2.7 years ("$s^2$" $= 2.7^2 - 0^2 \Rightarrow$ 
"$s$" $= 2.7$ years).\\ 

The total proper time on the space 
ships, between the event that clock $B$ left Earth and 
the event that clock $C$ returned to Earth, is 
$2.3 + 2.7 = 5$ years. This should be compared with the 
spacetime distance between the event that the space ship 
leaves Earth and the event that it comes back again, which 
was measured by the clock $A$ to be 9 years ("$s^2$" $= 9^2 - 
0^2 \Rightarrow$ "$s$" $= 9$ years). The proper time 
on Earth between the two events is thus 9 years. The 
proper time on Earth hence becomes longer than the 
total proper time on the space ships.\\ 
	
In the spacetime interpretation, the twin paradox thus 
arises as a \textit{geometric effect} in spacetime. Spacetime is 
not an Euclidean space, but distances are instead given by 
"$s^2$" $= t^2 - x^2$. This distance measure relates events 
in spacetime to one another, in an analogous way to how 
the distance measure given by the Pythagorean theorem 
relates points in the ordinary space to one another. A 
distance in spacetime does not only involve a distance 
in space, but is a mixture of a distance in space and a 
distance in time. The "triangle inequality in spacetime" 
mentioned above, describes how "proper time distances" 
between events in spacetime are related to one another. 
The spacetime distance between two events in spacetime 
is something that all observers or inertial reference 
systems agree on. In this case, there is no relativity. 
The spacetime distance measure dictates how "far" it is 
between each pair of events in spacetime.\\ 

What can be difficult, is to imagine how spacetime looks 
and hangs together as a whole. Just because we know how 
far it is between all events in spacetime, does not mean 
that we have an intuitive or clear picture of how spacetime 
globally seen looks.\\ 

Imagine the time before humans knew 
that the Earth was round and we thought that we lived in a 
world flat as a pancake. Let us imagine that our ancestors 
discovered that the distance between two points in their 
(and our) flat world was not given by the Pythagorean
theorem, but by a different distance measure. Suppose that 
our ancestors could mathematically describe this distance 
measure, and understood that they did not live in a flat 
world after all. Even if they themselves did not realize 
it, we know that it must be the distance measure for a 
spherical surface that our ancestors here had discovered. 
But as long as they do not understand this, the overall 
picture of the world they live in, will be hard for them 
to imagine and to get an intuitive feeling for.\\ 

A number 
of geometric effects could have arisen for these ancestors, 
that would have had to do with the fact that they, just 
like us, lived on a spherical surface and not a flat 
surface. E.g., two twins going in different directions from 
a point on the Earth's surface (e.g., the north pole), will 
eventually meet again at the point on the opposite side of 
the Earth's surface (the south pole); or that the sum of 
the angles of a triangle is not 180 degrees, but always a 
greater number.\\ 

There is however a flaw in this analogy, which has to do 
with the difference between their situation, which is 
being confronted by a spherical distance measure in a flat 
world, and our situation, which is being confronted by a 
non-Euclidean distance measure in an Euclidean world. For 
even if they lived in a flat world on the surface of Earth, 
their world was after all three-dimensional. They therefore 
knew what a spherical-shaped surface was, e.g., by
studying an orange, or a round stone. But neither we, nor 
our ancestors, have something in our everyday world, which 
has an equally directly obvious and intuitive
non-Euclidean geometry like the one that spacetime has. 
Thus, we do not really have any equally direct or obvious 
everyday experiences, which can give us an intuitive
feeling for what spacetime is and looks as a whole. 
Moreover, as we have discussed above, we humans are
probably equipped with a Newtonian approach to and 
notion of the world. We are therefore not familiar 
with looking at space and time in the manner that the 
spacetime interpretation tells us that we should. We 
often have to be satisfied with using mathematical 
descriptions, analogies, abstract concepts and images, 
to imagine what spacetime is and what it looks like.\\

But by accepting a spacetime interpretation, Einstein 
could with the general theory of relativity also come 
up with the idea that the geometry of spacetime could 
be \textit{curved}, analogous with how the geometry of 
our flat two-dimensional world on the surface of Earth is 
really curved as a two-dimensional spherical surface. 
By doing so, Einstein could explain acceleration and 
gravitation as two sides of the same coin, i.e., as 
effects of curvatures in spacetime. But this leads us 
into the general theory of relativity, which is not 
the subject of this paper.


\begin{thebibliography}{}

\bibitem[\protect\citeauthoryear{Bell}{Bell}{1993}]{bell93}
Bell, J.~S. (1993).
\newblock {Speakable and unspeakable in quantum mechanics}.
\newblock Cambridge university press.

\bibitem[\protect\citeauthoryear{Calder}{Calder}{1979}]{calder79}
Calder, N. (1979).
\newblock Einstein's universe.
\newblock {Book}, {176 pages}.

\bibitem[\protect\citeauthoryear{Einstein}{Einstein}{1905}]{einstein05}
Einstein, A. (1905).
\newblock Zur {E}lektrodynamik bewegter {K}\"orper.
\newblock {\em Annalen der Physik}, {\em 17}, 891--921.

\bibitem[\protect\citeauthoryear{Einstein}{Einstein}{1917}]{einstein17}
Einstein, A. (1917).
\newblock \"Uber die spezielle und die allgemeine Relativit\"atstheorie.
\newblock {Book}, {112 pages}.

\bibitem[\protect\citeauthoryear{Einstein}{Einstein}{1931}]{einstein31}
Einstein, A. (1931).
\newblock Mein Weltbild.
\newblock {Book}, {220 pages}.

\bibitem[\protect\citeauthoryear{Einstein}{Einstein}{1953}]{einstein53}
Einstein, A. (1953).
\newblock Aus Meinen Sp\"aten Jahren.
\newblock {Deutsche Verlags-Anstalt Stuttgart}, {278 pages}.

\bibitem[\protect\citeauthoryear{Einstein}{Einstein}{1951}]{Einstein51}
Einstein, A. (1951).
\newblock {Albert Einstein: Philosopher-Scientist}.
\newblock Tudor publishing company.

\bibitem[\protect\citeauthoryear{Epstein}{Epstein}{1985}]{epstein85}
Epstein, L. C. (1985).
\newblock Relativity visualized.
\newblock {Insight Press (San Francisco)}.

\bibitem[\protect\citeauthoryear{Levenson}{Levenson}{1996}]{levenson96}
Levenson, T. (1996).
\newblock Einstein revealed (TV).
\newblock {Director: Peter Jones}.

\bibitem[\protect\citeauthoryear{Schutz}{Schutz}{1995}]{schutz95}
Schutz, B. F. (1995).
\newblock A first course in general relativity 
\newblock {Book}, {ISBN 0-521-27703-5}.

\bibitem[\protect\citeauthoryear{Woodwardetal}{Woodwardetal}{1998}]{woodwardetal98}
Woodward, J. F. and Mahood, T. (1998).
\newblock What is the Cause of Inertia?
\newblock {\em Foundations of Physics}, {\em Vol. 29}, 
{\em No. 6}, {\em 1999}, pages 899--930.

\end{thebibliography}

\end{document}